\documentclass[twocolumn,showpacs,preprintnumbers,amsmath,amssymb,nofootinbib]{revtex4-1}

\usepackage[utf8]{inputenc}
\usepackage{graphicx}
\usepackage{dcolumn}
\usepackage{float}
\usepackage{color}
\usepackage{bm}

\newcommand{\bea}{\begin{equation}\begin{array}{c}}
\newcommand{\eea}{\end{array}\end{equation}}
\newcommand{\ea}{\end{array}}

\newcommand{\be}{\begin{equation}}
\newcommand{\ee}{\end{equation}}
\newcommand{\bad}{\begin{array}{ccc}}

\newcommand{\ba}{\begin{array}{c}}
\newcommand{\diag}{{\rm diag}}

\begin{document}
\title{{\color{blue}\bf A see-saw scenario of an $A_4$ flavour symmetric\\ standard model}}
\author{Dinh Nguyen Dinh$^{1,2}$}
\email{dndinh AT iop.vast.ac.vn}
\author{Nguyen Anh Ky$^{3,4}$} 
\email{anhky AT iop.vast.ac.vn}
\email{\textit{\\Permanent address: Institute of physics, VAST, Hanoi.}}
\author{Phi Quang V$\breve{\mbox{a}}$n$^{1}$}
\email{pqvan AT iop.vast.ac.vn}
\author{Nguyen Thi H$\grave{\hat{\mbox{o}}}$ng V$\hat{\mbox{a}}$n$^{1,5}$}
\email{nhvan AT iop.vast.ac.vn}
\affiliation{\vspace*{2mm}
$^1$\textit{Mathematical and high energy physics group},\\ 
\textit{Institute of physics},\\ 
\textit{Vietnam academy of science and technology},\\ 
\textit{10 Dao Tan, Ba Dinh, Hanoi, Viet Nam.} 
\vspace*{2mm}\\
$^2$ \textit{Department of Physics},\\ 
{\it University of Virginia,}\\ 
{\it  Charlottesville, VA 22904-4714, USA.}\vspace*{2mm}\\
$^3$\textit{Duy Tan university},\\
\textit{K7/25 Quang Trung street, Hai Chau, Da Nang, Viet Nam.}\vspace*{2mm}\\
$^4$\textit{Laboratory of high energy physics and cosmology},\\ 
\textit{Faculty of physics},\\
\textit{VNU university of science},\\ 
\textit{334 Nguyen Trai, Thanh Xuan, Hanoi, Viet Nam.}\vspace*{2mm} \\ 
$^5$\textit{Institute for interdisciplinary research in science and education},\\ 
\it{ICISE, Quy Nhon, Vietnam.}}
%

%
%


\date{\today}

\begin{abstract}
A see-saw scenario for an $A_4$ flavour symmetric standard model is presented. 
The latter, compared with the standard model, has an extended field content 
adopting now an additional $A_4$ symmetry structure (along with the standard model symmetry). As before, the see-saw mechanism can be realized in several models of 
different types depending on different ways of neutrino mass generation corresponding 
to the introduction of new (heavy in general) fields with different symmetry structures. 
In the present paper, a general description of all these see-saw types is made with a 
more detailed investigation on type-I models, while for type-II and type-III models a similar strategy can be followed. As within the original see-saw mechanism, the symmetry structure of the standard model fields decides the number and the symmetry structure of 
the new fields. In a model considered here, the scalar sector consists of 
three standard-model-Higgs-like iso-doublets ($SU_L(2)$-doublets) forming 
together  an $A_4$-triplet, and three iso-singlets transforming as three 
singlets (1,$1^{'}$ and $1^{''}$) of $A_4$. 
In the lepton sector, the three left-handed lepton iso-doublets form an 
$A_4$-triplet, while the three right-handed charged leptons are either 
$A_4$-singlets in one version of the model, or components of an $A_4$-triplet 
in another version. To generate neutrino masses through, say, the type-I see-saw 
mechanism, it is natural to add four right-handed neutrino multiplets, including 
one $A_4$-triplet and three $A_4$-singlets. For an interpretation, the model is 
applied to deriving some physics quantities such as neutrinoless double beta decay 
effective mass $|\langle m_{ee}\rangle|$, CP violation phase $\delta_{CP}$ and 
Jarlskog parameter $J_{CP}$, which can be verified experimentally.  
\end{abstract}

\maketitle


\section{\label{sec:level1}Introduction}
Although the standard model (SM) \cite{sm1,sm2,qft,quangyem} has proved 
to be a very successful model of elementary particles and their interactions, 
especially after the discovery of the Brout-Englert-Higgs boson, or, shortly, 
the Higgs boson, by the LHC collaborations ATLAS and CMS 
\cite{Aad:2012tfa,Chatrchyan:2012ufa} (see also \cite{Ky:2015eka} for 
a review), it, however, cannot solve a number of problems in particle 
physics and astrophysics. One of such problems is that of neutrino masses 
and mixing which is an experimental fact. These problems often require an 
extension of the SM. Among many extended, or say, beyond standard model (BSM), 
models, suggested, the models based on a flavour symmetry have attracted  
much interest for over one decade. In these models, the original SM fields, 
including neutrinos, along with new fields which may be added, are assumed to 
adopt a flavour symmetry (transformation) structure.\\

In the SM, neutrinos are massless and not mixing, but experimental results on 
neutrino oscillations 
\cite{Fukuda:1998tw,Fukuda:1998ub,Fukuda:1998mi,Ahmad:2001an,Ahmad:2002jz,Ahmad:2002ka} 
have shown that neutrinos are massive and mixing. A flavour neutrino, thus, is a mixture 
of light neutrinos $\nu_i$ (where $i=1,2,3$, for a three-neutrino model) with masses, 
say $m_i$, expected so far to be smaller than $1$ eV (see, e.g., \cite{Olive:2016xmw}). 
The (three-neutrino) mixing matrix in the canonical form, known as 
Pontecorvo-Maki-Nakagawa-Sakata (PMNS) neutrino mixing matrix, and parametrized by 
three mixing angles, $\theta_{12}$, $\theta_{23}$, $\theta_{13}$, one Dirac phase 
$\delta$ and, if neutrinos are Majorana particles, two more Majorana phases, 
$\alpha_{21}$ and $\alpha_{31}$, can be written in the form 
\begin{equation}
\label{PMNS0}
U_{PMNS}=U\times P,
\end{equation}
where
\begin{equation}
 P={\rm diag}(e^{i\alpha_{21}/2},e^{i\alpha_{31}/2},1),
\label{p}
\end{equation}
and
\begin{widetext}
\begin{equation}
\label{PMNS}
U=
\left(
\begin{array}{ccc}
 c_{12}c_{13} & s_{12}c_{13} & s_{13} e^{-i \delta} \\
 -c_{23}s_{12}-s_{13}s_{23}c_{12} e^{i \delta} & c_{23}c_{12}-s_{13}s_{23}s_{12} 
 e^{i \delta} & s_{23}c_{13} \\
 s_{23}s_{12}-s_{13}c_{23}c_{12} e^{i \delta} & -s_{23}c_{12}-s_{13}c_{23}s_{12} 
 e^{i \delta} & c_{23}c_{13}
\end{array}
\right),
\end{equation}
\end{widetext}
with the notations 
$s_{ij}=\sin \theta_{ij}$ and $c_{ij}=\cos\theta_{ij}$, $1\leq i<j\leq 3$, used. 
Here, the mixing angles and the phases are taken within the following ranges: 
$\theta_{ij}\in [0,\pi/2]$, $\delta \in [0,2\pi]$, $\alpha_{j1} \in [0,2\pi]$, 
$j=2,3$. As the Dirac phase $\delta$ is related to the CP violation (CPV) 
phenomenon, it is also called the CPV phase and denoted as $\delta_{CP}$ (but 
below the short notation $\delta$ is more frequently used). The 
best fit values of the neutrino oscillation angles, given in 
\cite{Olive:2016xmw,Capozzi:2017ipn, Esteban:2016qun}, are close to the tri-bi-maximal (TBM) 
mixing, where 
$\sin^2\theta_{12}^{\tiny{TBM}}=1/3$, $\sin^2\theta_{23}^{TBM}=1/2$ 
and $\sin^2\theta_{13}^{TBM}=0$ 
\cite{Harrison:2002er}. The most recent neutrino oscillation experimental data 
\cite{Olive:2016xmw,Capozzi:2017ipn, Esteban:2016qun}, summarized in Table \ref{Synopsis} for 
both the normal ordering (NO) and the inverse ordering (IO) of neutrino masses, 
has shown, however, a deviation of the PMNS matrix from 
the TBM form. There have been many attempts, including an assumption of a flavour 
symmetry (see more details below), to explain this phenomenon.\\
%
%
\begin{widetext}
\begin{center}
\begin{table}[H]
\begin{center}
\begin{tabular}{|l|c|c|c|c|}
\hline
\quad\quad\quad~ Parameter & Best fit & $1\sigma$ range & $2\sigma$ range & $3\sigma$ range\\
\hline
$\Delta m_{21}^2/10^{-5}~\mathrm{eV}^2 $ (NO or IO) & 7.54 & 7.32 -- 7.80 & 7.15 -- 8.00 & 6.99 -- 8.18 \\
\hline
$\sin^2 \theta_{12}/10^{-1}$ (NO or IO) & 3.08 & 2.91 -- 3.25 & 2.75 -- 3.42 & 2.59 -- 3.59 \\
\hline
$\Delta m_{31}^2/10^{-3}~\mathrm{eV}^2 $ (NO) & 2.47 & 2.41 -- 2.53 & 2.34 -- 2.59 & 2.27 -- 2.65 \\
$|\Delta m_{32}^2|/10^{-3}~\mathrm{eV}^2 $ (IO) & 2.42 & 2.36 -- 2.48 & 2.29 -- 2.55 & 2.23 -- 2.61 \\
\hline
$\sin^2 \theta_{13}/10^{-2}$ (NO) & 2.34 & 2.15 -- 2.54 & 1.95 -- 2.74 & 1.76 -- 2.95 \\
$\sin^2 \theta_{13}/10^{-2}$ (IO) & 2.40 & 2.18 -- 2.59 & 1.98 -- 2.79 & 1.78 -- 2.98 \\
\hline
$\sin^2 \theta_{23}/10^{-1}$ (NO) & 4.37 & 4.14 -- 4.70 & 3.93 -- 5.52 & 3.74 -- 6.26 \\
$\sin^2 \theta_{23}/10^{-1}$ (IO) & 4.55 & 4.24 -- 5.94 & 4.00 -- 6.20 & 3.80 -- 6.41\\
\hline
\end{tabular}
\end{center}
\caption{{\small \label{Synopsis}  The best-fit values and $n\sigma$ allowed ranges of the 3-neutrino oscillation
parameters for a normal neutrino mass ordering (NO) and an inverse neutrino mass ordering (IO) 
\cite{Olive:2016xmw,Capozzi:2017ipn, Esteban:2016qun}.}}
\end{table}
\end{center}
\end{widetext}
%

Besides the parameters in the PMNS mixing matrix, neutrino oscillation experiments 
also allow us to determine the neutrino squared mass differences $\Delta m_{21}^2$ 
and $\Delta m_{31}^2$. Although the absolute neutrino masses have not yet been known  
but it is believed that they are very tiny, less than 1 eV, as said above. Therefore, 
one needs to understand why the neutrinos, compared with the charged leptons, are so 
light, and find a mechanism for generation of such small masses. A very popular 
mechanism of this kind is called the see-saw mechanism which can generate a small 
neutrino mass due to an introduction of a larger scale which could be a mass of a 
heavy, compared with neutrinos, BSM particle. The see-saw mechanism, appearing first 
as type I in \cite{Minkowski:1977sc,GellMann:1980vs,Mohapatra:1979ia,Schechter:1980gr,
Schechter:1981cv}, has been originally applied to the SM (see 
\cite{Bilen,Mohapatra:1998rq} for later developments and a more complete presentation) 
but a natural question arising here is whether this mechanism can be incorporated in 
a model with an additional symmetry. This idea has  
attracted interest of other authors investigating different models, including those 
with a flavour symmetry adopted (see, for example, \cite{Caldwell:1993kn, Petcov:1993rk}).
\\ 

Various models with flavour symmetries have been introduced to explain the mass spectrum 
and the mixing matrix of the quarks and leptons. Among them, the models constructed with 
non-Abelian discrete flavour symmetries added are investigated quite intensively 
(see \cite{King:2014nza,Ishimori:2010au,Altarelli:2010gt} for a review). Especially, 
a number of models that are not only able to give a TBM mixing as well as adjustable 
to fit the data of the observed neutrino oscillations but also interesting due to their 
simplicity, are based on the $A_4$ flavour symmetry 
\cite{Ishimori:2010au,Altarelli:2010gt, Ma:2001dn, Ma:2002iq, Babu:2002dz, 
Ma:2004zv,Altarelli:2009kr,Parattu:2010cy, 
King:2011ab,Altarelli:2012bn,Altarelli:2012ss,Ferreira:2013oga,Barry:2010zk,Ahn:2012tv,
Felipe:2013vwa,Hernandez:2013dta,Varzielas:2015joa,Hung:2015nva,Ky:2016rzl}. 
There are many models based on other flavour symmetries but they are beyond the scope 
of this paper. 
\\

In the present work, a ``naturally" extended see-saw version of the SM adopting an $A_4$ 
flavour symmetry is suggested and considered. In this model the three generations of  the 
left-handed leptons are grouped in a triplet of $A_4$, while the three right-handed charged 
leptons either transform under $A_4$ as its 1, $1^{'}$ and $1^{''}$ singlets (but there is 
another version, in which the  right-handed charged leptons form an $A_4$ triplet). The 
scalar sector is extended with the SM Higgs boson acquiring now an $A_4$ triplet structure 
and three additional scalar iso-singlets transforming as 
1, $1^{'}$ and $1^{''}$ singlets of $A_4$. 
Light neutrino masses can be generated via the type I see-saw mechanism by introducing 
four right-handed neutrinos, which are one triplet and three singlets $(1,1',1'')$ under 
the transformation of the $A_4$ group. This model is a multiple model of several copies 
corresponding to different representations of $A_4$. To our knowledge, this approach is 
done for the first time here, and it reminds of the 
co-variant diagram approach in supersymmetry models \cite{Gates:1983nr,Grisaru:1984ja,
Galperin:1987vx,Ky:1988yn}.  \\

In Section 2 we present the main idea of the original see-saw mechanism applied to the SM 
and extend it to the case with a flavour symmetry involved. Sections 3 and 4 are devoted 
to a more detailed investigation on a type I see-saw mechanism within the newly suggested  
$A_4$-flavour symmetric SM which is checked by calculating some physics quantities for 
neutrino masses and mixing and comparing them with the current experimental data. 
%
\\
Section 5 is designed for some comments and conclusion. The bibliography of the 
cited works may be still far from being complete but we hope it gives a general 
view on the development of the topic considered 
here. Before going to physics details in the sections following, the reader can 
have a look at the appendix for notations and a quick review on  the basic 
elements of $A_4$ representations.   

\section{See-saw mechanism for an $A_4$-flavour symmetric standard model}
Neutrinos in the standard model (SM) are massless but, as said above, 
experimental results 
\cite{Fukuda:1998tw,Fukuda:1998ub,Fukuda:1998mi,Ahmad:2001an,Ahmad:2002jz,Ahmad:2002ka} 
have shown that they have masses though very tiny. The see-saw mechanism 
\cite{Bilen,Mohapatra:1998rq} is an attempt to explain 
the neutrino mass smallness. Let us first recall the see-saw mechanism applied to the 
(original) SM. Generally speaking, this mechanism imposes an extension on 
the SM by adding in different ways new (heavy in general) fields in order to generate 
(small effective) masses of neutrinos (due to big masses of the new fields). These ways 
of neutrino mass generation are called see-saw models referred below to as classical or 
pre-flavour symmetric see-saw models.

\subsection{Pre-flavour-symmetric see-saw models}
\hspace{0.5cm}
There are three types of classical see-saw models: I, II and III, 
corresponding to three ways of neutrino mass generation, requiring 
an introduction of three kinds of new fields which in general are 
heavy. The see-saw I generates neutrino masses with the introduction 
of a lepton iso-singlet (a right-handed neutrino), while the 
generation of neutrino masses via the see-saw II and III requires 
respectively a new scalar iso-triplet and a new lepton iso-triplet 
to be introduced. Let us look at a closer distance how they work. 
\\

A neutrino mass could be of Dirac- or Majorana type. 
A general Lagrangian neutrino mass term, denoted below 
by $\mathcal{L}_{m_\nu}$, has the form 
\begin{equation}
\mathcal{L}_{m_\nu}= -\frac{1}{2}\overline{n}_LM_{\nu}(n_L)^c+h.c.,
\end{equation}
where
\begin{equation}
\label{Eq42}
n_L=
\begin{pmatrix}
\nu_L\\
(N_R)^c\\
\end{pmatrix}
\end{equation}
is a neutrino pattern, and
\begin{equation}
M_{\nu}=
\left( 
\begin{array}{cc}
M_L & M_D\\
(M_D)^T & M_R\\
\end{array}
\right)
\label{ssmass}
\end{equation}
is a neutrino mass matrix ($M_L$ is a $3\times 3$ matrix for the case of three left-handed neutrinos $\nu_L$, 
and $M_R$ is an $n\times n$ matrix for the case of $n$ right-handed neutrinos $N_R$, 
then $M_D$ is a $3\times n$ matrix).
If $M_L \ll M_R$, the matrix \eqref{ssmass} can be approximately diagonalized 
in the block form 
\cite{Bilen,Mohapatra:1998rq} 
\begin{equation}
M_{\nu}=
\left( 
\begin{array}{cc}
M_1 & 0\\
0 & M_2\\
\end{array}
\right),
\label{ssmd}
\end{equation}
where $M_1$ and $M_2$,
\begin{equation}
\begin{cases} 
M_1\approx M_L-(M_D)^T (M_R)^{-1} M_D, \\
M_2\approx M_R,
\end{cases}
\label{seesaw}
\end{equation}
are complex matrices and they are the mass matrices of the light- and 
heavy neutrinos, respectively. The light neutrinos, which are (mostly) left-handed 
and can interact with ordinary SM fields, are called active neutrinos, while the 
heavy, and right-handed here, neutrinos, which, in many models, interact very 
weakly or do not interact at all with the SM fields, are referred to as sterile 
neutrinos. The constraint \eqref{seesaw} means that the masses of the light 
neutrinos, upto a small amount ($M_L$), are inversely proportional to the masses 
of the heavy neutrinos. 
This is the general spirit, thus the name, of the see-saw mechanics for generation  
of a small neutrino mass with a mass scale $M_R$ taking usually a value very large, 
between $10^9$ GeV and the GUT scale, around $10^{15}$ GeV \cite{Drewes:2013gca}, 
but in some see-saw- and other models, $M_R$ can varies within a wider range, 
even it could take a value at few keV's or less (see, for example, 
\cite{Drewes:2013gca,Dinh:2006ia,Ky:2005yq,Merle:2013gea,Adhikari:2016bei}).\\

 A large $M_R$ can be generated via a newly-introduced heavy field. 
It can be done in different ways, corresponding to different see-saw models, 
depending on how a coupling of a neutrino to a scalar field is constructed, 
as we still assume that neutrinos can acquire mass via interacting somehow 
with a scalar field. 
Since, in the SM, a neutrino belongs to an iso-doublet \textbf{2} and the 
Higgs field forms another iso-doublet \textbf{2}, there are (at the tree level, 
as illustrated in Figs. \ref{eff-mass}--\ref{ss1-3} below) three possible ways 
of their coupling due to the tensor product decomposition 
$\textbf{2}\times \textbf{2} =\textbf{1} + \textbf{3}$: two ways 
in which neutrino and Higgs are coupled at a vertex to a fermion iso-singlet or 
a fermion iso-(anti)triplet, and one way, two ways at first sight, in which 
neutrino is coupled to Higgs via a scalar iso-(anti)triplet (see more details below). 
The see-saw models constructed with neutrino-Higgs coupled to a fermion iso-singlet, 
a scalar iso-(anti)triplet and a fermion iso-(anti)triplet are called of type I, 
II and III, respectively. \\

The see-saw mechanism can be linked to the high-dimension effective operator 
approach such as Weinberg's 5-dimension one in which the neutrino (effective) 
mass can originate from the Lagrangian effective term $\mathcal{L}_\nu^{eff}$, 
\begin{equation}
\mathcal{L}_{\nu}^{eff}=\frac{\lambda_{\nu}}{\Lambda}\left(\overline{f}_{lL} 
H \right) \left(H^{\dagger} f_{lL}^c \right),
\end{equation}
where $\Lambda$ is a characteristic scale, 
\begin{equation}
f_l = \left(
\begin{array}{c}
\nu_{lL} \\
\l_L
\end{array}
\right),  \hspace{2mm} l=e, \mu, \tau, 
\end{equation}
and
\begin{equation}
H = \left(
\begin{array}{c}
H^0\\
H^-
\end{array}
\right).
\end{equation}
This effective process\footnote{It may have other substructures if we consider other 
models such as loop models, supersymmetry models, extra-dimension models, etc., which 
are not a subject of the present paper.}, illustrated by Fig. \ref{eff-mass}, can be 
broken down into subprocesses such as the ones illustrated in Fig. \ref{ss1-3}, where 
the first diagram illustrates see-saw models of type I and III, which respectively 
require the introduction of new fields, say $N$ and $\Sigma$, being respectively a 
fermion iso-singlet and a fermion iso-(anti)triplet, while the second diagram illustrates 
a see-saw model of type II, which requires the introduction of a new field $\Delta$, being 
a scalar iso-(anti)triplet. Mathematically, one can ask a question if it is possible to introduce a model with $\Delta$ 
replaced by a scalar iso-singlet, say $S$, but the latter is not relevant to describe a neutrino mass term (because it is impossible to contact an antisymmetrized $S$ (in iso-indices) with a symmetrized fermion couple). In these figures and all figures bellow, the letters $\nu$ and $H$ 
symbolically denote the fields ``neutrino" and ``Higgs" (or their anti fields), respectively.  
\begin{widetext}
\begin{center}
\begin{figure}[H] 
\begin{center}
\begin{tabular}{c}
\includegraphics[scale=0.3]{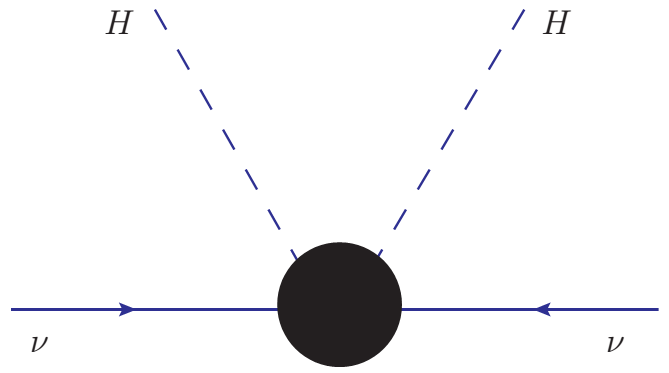} 
\end{tabular}
\caption{\small Neutrino effective mass.}
\label{eff-mass}
\end{center}
\end{figure}
\begin{figure}[H]
\begin{center}
\begin{tabular}{cc}
\includegraphics[scale=0.3]{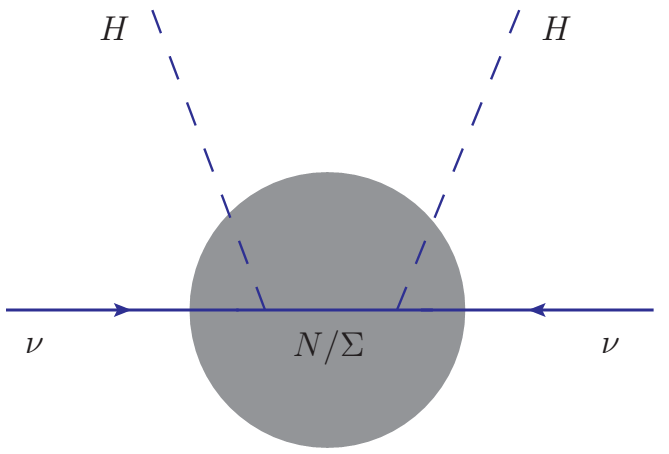} ~~~~
&
~~~~ \includegraphics[scale=0.3]{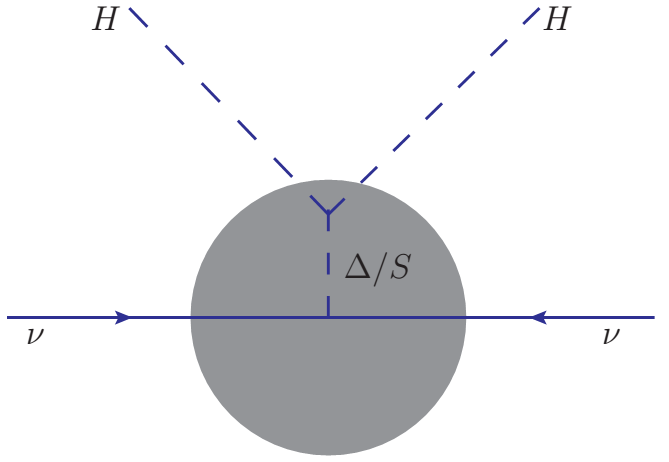}
\end{tabular}
\caption{\small See-saw models.}
\label{ss1-3}
\end{center}
\end{figure}
\end{center}
\end{widetext}
~\\
When the fields transform under an additional symmetry group, such as an $A_4$ one, 
each of these (sub)processes itself will be further broken down to sub-sub-processes 
according to their symmetry structure. It is the spirit of the see-saw mechanism in 
a new scenario with an additionally introduced symmetry.  
%
%
\subsection{$A_4$-flavour symmetric see-saw models}
\begin{figure}[H]
\begin{center}
\begin{tabular}{c}
\includegraphics[scale=0.2]{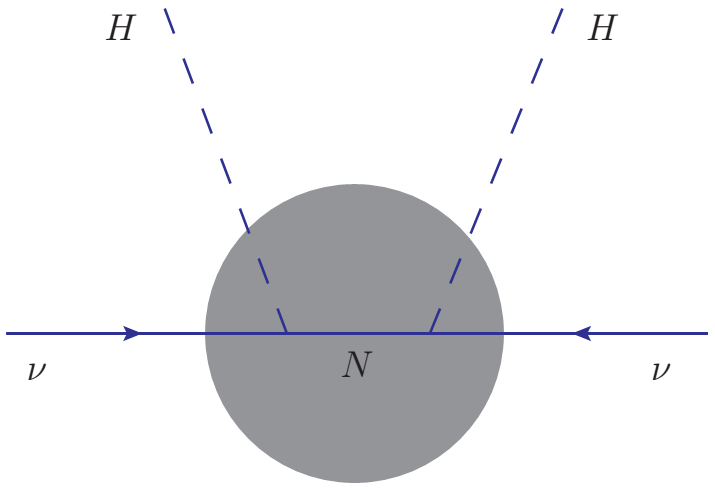} 
\end{tabular}
\caption{\small Type-I see-saw neutrino effective mass.}
\label{A4-eff-mass}
\end{center}
\end{figure}

\hspace{0.5cm}
There are also three types of see-saw models for this case corresponding to the 
above-described type I, II and III see-saw models (for some versions, see, 
for example, \cite{Barry:2010zk,Felipe:2013vwa} and references therein). 
Now, each of the processes in Fig. \ref{ss1-3} 
becomes an effective process containing other (sub)processes. Let us consider an 
effective process, illustrated in Fig. \ref{A4-eff-mass}, corresponding to the type I 
see-saw one, where $N$ is a fermion iso-singlet (or a set of fermion iso-singlets). 
This process in turn represents, accordingly to the rule \eqref{tensor33}--Appendix, 
a sum of different sub-processes (see Fig. \ref{A4-ss1} below). We will see this model 
in more details 
next section. The type III model (i.e., the model corresponding to the type III see-saw 
model) has a similar structure, where the fermion iso-singlets $N$'s are replaced by 
fermion iso-triplets $\Sigma$'s being also a triplet and singlets $1$, $1^{'}$ and 
$1^{''}$ of $A_4$. 

\section{$A_4$-flavour symmetric see-saw-I model}
We present in this section an $A_4$-flavour symmetric extension of a type-I 
see-saw model, while the see-saw models of type-II and type-III can be 
investigated in separate works.  
There could be different versions of this extended model but 
the one given in Tab. \ref{ssI1} requires 
a minimal and ``natural" extension, thus, this model can be referred to as the 
minimally extended model of the type-I see-saw model, or, just, the minimal 
model, for short. 
\begin{table}[H]
\begin{center}
\begin{tabular}{|l||c|c||c|c|c|c||c|c|c|c|c|}
\hline
           & $\ell_{L}$ & $\ell_{Ri}$ & $\Phi_{h}$  & $\Phi_S$ & $\Phi_{S^{'}}$&$\Phi_{S^{''}}$ &  $N_T$ &  $N_S$  & $N_{S'}$    
& $N_{S''}$     \\  \hline
Spin & 1/2 & 1/2 & 0 & 0 & 0 & 0 & 1/2 & 1/2 & 1/2 & 1/2          \\ \hline
 $SU(2)_L$ &    2     &    1, 1, 1     &   2  &  1 & 1  & 1 &  1     &       1      &    1        &  1       \\
 \hline
 $A_4$        &    3     &  $1, 1^{'}, 1^{''}$    &      3  & 1 & $1^{'}$& $1^{''}$&  3      &     1      
& $1^{'}$     & $1^{''}$    \\  \hline
\end{tabular}
\caption{An $A_4$-flavour symmetric extended standard model}
\label{ssI1}

\end{center}
\end{table}
The group transformation nature of the fields in the 
considered model is shown on the table, where 
$\ell_{L}=(\ell_{L1},\ell_{L2},\ell_{L3})$ and $\ell_{Ri}$, $i=1,2,3$, are the three generations of the 
left-handed- and the right-handed charged leptons, respectively. The three iso-doublets $\ell_{Li}$ form together an $A_4$-triplet, while the iso-singlets $\ell_{Ri}$ are also $A_4$-singlets, 1, $1'$ and $1''$ (we can consider another version in which $\ell_{Ri}$ 
are components of an $A_4$-triplet). We note that the states $\ell_i$ ($\ell_{Li}$ and 
$\ell_{Ri}$) in general may not coincide with the states $l=e,\mu,\tau$, 
to which we need to make a rotation given in the subsection following. The new fermions 
$N_T$, $N_S$, $N_{S'}$ and $N_{S''}$, being neutral fields and iso-singlets, are  
an $A_4$-triplet and three $A_4$-singlets 1, $1'$ and $1"$, respectively.\\

 The scalar field  
\begin{equation}
\Phi_h^T=(\phi_{h1}, \phi_{h2}, \phi_{h3}),
\end{equation}
is an $A_4$-triplet composed of three SM-like Higgs fields $\phi_{hi}$
being iso-doublets,
\begin{equation}
\phi_{hi} =
\left(
\begin{array}{c}
\varphi_i^+   \\[2mm]
\varphi_i^0
\end{array}
\right),~~ i=1,2,3.
\end{equation}
The vacuum expectation value (composed of those 
of the neutral components) of the scalar field $\Phi^0_h$ is denoted as 
\begin{equation}
\langle \Phi^0_h \rangle ^T= \left(\langle\varphi_1^0\rangle,\langle\varphi_2^0\rangle,
\langle\varphi_3^0\rangle\right):={1\over \sqrt{2}}(v_1,v_2,v_3). 
\label{phi-vev}
\end{equation}
The structure of this vacuum expectation value (VEV), as well as the VEV's of other scalar fields, can be fixed when 
the potential of the scalar sector is considered.  The fields $\Phi_S$, $\Phi_{S^{'}}$ and $\Phi_{S^{''}}$ are iso-singlet and $A_4$-singlet scalars with  
\begin{equation}
\langle \Phi_S \rangle =\sigma_1,~~\langle \Phi_{S^{'}} \rangle =\sigma_2,~~\langle \Phi_{S^{''}} 
\rangle =\sigma_3, 
\end{equation}
denoting their VEV's. These scalar fields can be re-expressed in terms of quantum fields $h_i$, $\eta_i$ and $\xi_i$ with zero VEV's by shifting 
\begin{widetext}
\begin{equation}
\phi_{h1} =
\left(
\begin{array}{c}
\varphi_1^+   \\[2mm]
{1\over \sqrt{2}}(v_1 +h_1+i\eta_1)
\end{array}
\right),  \hspace{0.3cm}
\phi_{h2} =
\left(
\begin{array}{c}
\varphi_2^+   \\[2mm]
{1\over \sqrt{2}}(v_2 +h_2+i\eta_2)
\end{array}
\right),  \hspace{0.3cm}
\phi_{h3} =
\left(
\begin{array}{c}
\varphi_3^+   \\[2mm]
{1\over \sqrt{2}}(v_3 +h_3+i\eta_3)
\end{array}
\right),
\end{equation} 
\end{widetext}
and 
\begin{equation}
\Phi_S=\sigma_1+\xi_1, ~~ \Phi_{S'}=\sigma_2+\xi_2, ~~ \Phi_{S''}=\sigma_3+\xi_3.
\end{equation}

Below, we shall choose to consider the type I see-saw model given in 
Tab. \ref{ssI1} and illustrated in Fig. \ref{A4-ss1}.
\begin{widetext}
\begin{center}
\begin{figure}[H]
\begin{center}
\begin{tabular}{cccc}
\includegraphics[scale=0.3]{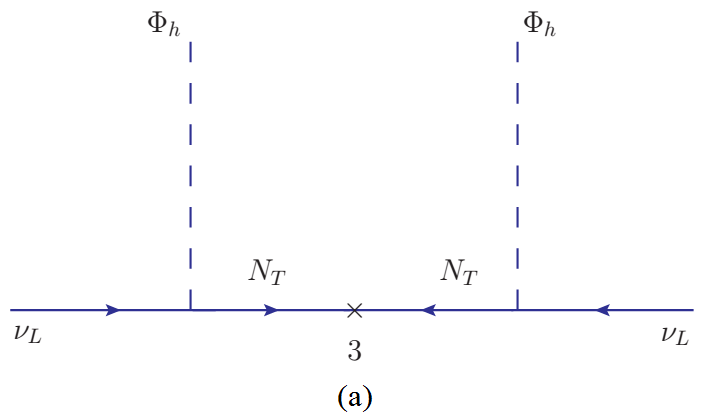} ~~
& ~~
\includegraphics[scale=0.3]{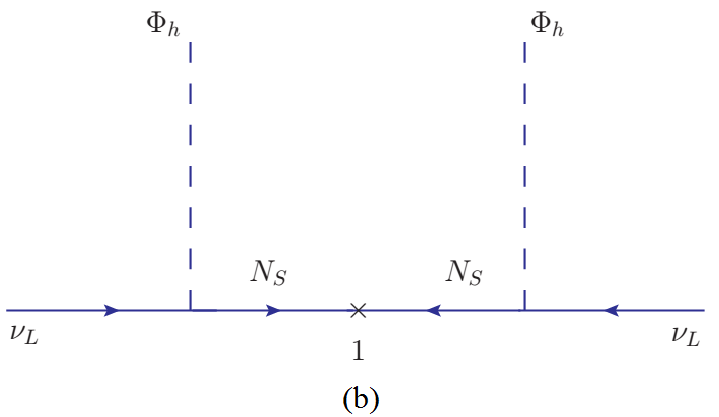}
\end{tabular}
\end{center}
\end{figure}
\begin{figure}[H]
\begin{center}
\begin{tabular}{cc}
\includegraphics[scale=0.3]{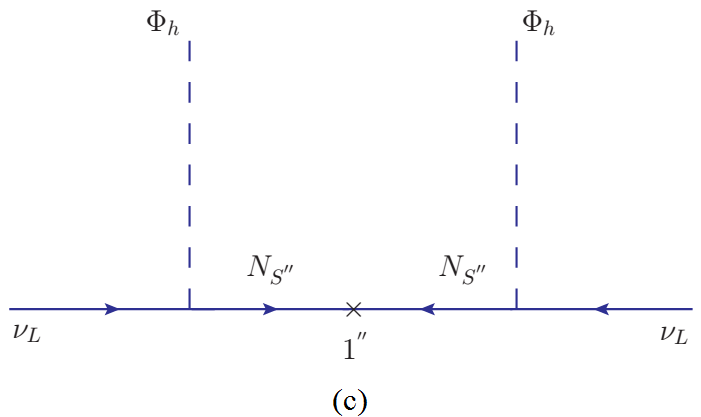} ~~
& ~~
\includegraphics[scale=0.3]{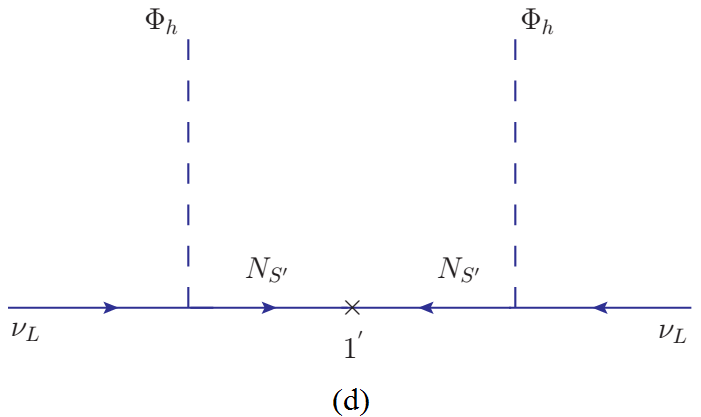}
\end{tabular}
\caption{An $A_4$-flavour-symmetric type-I see-saw model.}
\label{A4-ss1}
\end{center}
\end{figure}
\end{center}
\end{widetext}

Now, let us make a brief description of the scalar- and the lepton sector of the model 
(the quark sector not considered here requires a separate investigation). 
%

	\subsection{Scalar sector}
	\hspace{0.5cm}	This sector consists of four scalars which are an iso-doublet $A_4$-triplet $\Phi_h$, and three iso-singlets $\Phi_S$, $\Phi_{S^{'}}$ and $\Phi_{S^{''}}$ transforming as $A_4$-singlets 1, $1^{'}$ and $1^{''}$, respectively. A logical 
	model construction requires the symmetry breaking to follow this scheme: 
	\begin{equation}
	\begin{array}{c}
	A_4\otimes SU(2)_L \otimes U(1)_N\\[2mm]
	\downarrow \langle \mathcal{S} \rangle \\[2mm]
	SU(2)_L \otimes U(1)_Y\\[2mm]
	\downarrow \langle \Phi_h \rangle \\[2mm]
	U(1)_Q, \end{array}\label{sb}\end{equation}
	where $\mathcal{S}$ stands for the scalars $\Phi_S$, $\Phi_{S^{'}}$ 
	and $\Phi_{S^{''}}$.
	The additional scalar fields may play a similar role 
	as that of the scalar fields in a supersymmetric model \cite{Martin:1997ns} in compensating divergences of fermion fields, therefore, we may not need SUSY in building a finite field theory (to have a precise conclusion we must make a detailed analysis which is beyond the subject of this paper). \\

	Here, we briefly discuss the Higgs potential in the model. It has the general form
	\begin{equation}
	V = V (\Phi_h)+V(\Phi_h, \Phi_S, \Phi_{S^{'}}, \Phi_{S^{''}})+V(\Phi_S, \Phi_{S^{'}}, \Phi_{S^{''}}),
	\end{equation}
	with
\begin{widetext}	
	\begin{align}
	V(\Phi_h) =&~ \mu^2_0 (\Phi_h^{\dagger}\Phi_h)_{1}+\lambda_1 (\Phi_h^{\dagger}\Phi_h)^2_{1}
	+\lambda_2 (\Phi_h^{\dagger}\Phi_h)_{1^{'}} (\Phi_h^{\dagger}\Phi_h)_{1^{''}}    \nonumber   \\
	&+ \lambda_3 (\Phi_h^{\dagger}\Phi_h)_{3_s}(\Phi_h^{\dagger}\Phi_h)_{3_s}
	+\lambda_4 (\Phi_h^{\dagger}\Phi_h)_{3_a}(\Phi_h^{\dagger}\Phi_h)_{3_a}
	+\lambda_5 (\Phi_h^{\dagger}\Phi_h)_{3_s}(\Phi_h^{\dagger}\Phi_h)_{3_a}, 
	\label{VPhi} 
	\end{align}
	\begin{align}
	V(\Phi_h, \Phi_S, \Phi_{S^{'}}, \Phi_{S^{''}}) =&~ \gamma_1 (\Phi_h^{\dagger}\Phi_h)_{1} (\Phi_S^{\dagger}\Phi_S)_{1}
	+ \gamma_2 (\Phi_h^{\dagger}\Phi_h)_{1} (\Phi_{S'}^{\dagger}\Phi_{S^{''}})_{1}   \nonumber  \\
	&+\beta_1 (\Phi_h^{\dagger}\Phi_h)_{1^{'}} (\Phi_{S'}^{\dagger}\Phi_{S^{'}})_{1^{''}} +\beta_2 (\Phi_h^{\dagger}\Phi_h)_{1^{'}} (\Phi_{S''}^{\dagger}\Phi_S)_{1^{''}} 
	\nonumber \\
	&+\eta_1 (\Phi_h^{\dagger}\Phi_h)_{1^{''}} (\Phi_{S''}^{\dagger}\Phi_{S^{''}})_{1^{'}} +\eta_2 (\Phi_h^{\dagger}\Phi_h)_{1^{''}} (\Phi_{S'}^{\dagger}\Phi_S)_{1^{'}} +H.c.,
	\label{Vphi3s}
	\end{align}
	\begin{align}
	V(\Phi_S, \Phi_{S^{'}}, \Phi_{S^{''}}) =& ~\alpha_0 (\Phi_S^{\dagger}\Phi_S)_{1}+\alpha_1 (\Phi_S^{\dagger}\Phi_S)_{1}^2  \nonumber  \\
	&+\alpha_2 (\Phi_S^{\dagger}\Phi_S)_{1}(\Phi_{S'}^{\dagger}\Phi_{S^{''}})_{1}+\alpha_3 (\Phi_{S'}^{\dagger}\Phi_{S^{''}})_{1}   \nonumber  \\
	&+\alpha_4 (\Phi_{S'}^{\dagger}\Phi_{S^{''}})_{1} (\Phi_{S'}^{\dagger}\Phi_{S^{''}})_{1} +\alpha_5 (\Phi_{S'}^{\dagger}\Phi_{S^{'}})_{1^{''}} (\Phi_{S''}^{\dagger}\Phi_{S^{''}})_{1^{'}} 
	+H.c..
	\end{align}
\end{widetext}
	
	For a logical reason, we require that $A_4$ symmetry should be broken at an energy scale higher (or much higher) than the electroweak one. This symmerty breaking can be caused when some or all of the scalar fields $\Phi_S$, $\Phi_{S^{'}}$ and $\Phi_{S^{''}}$ acquire non-zero VEV's. Therefore, $\Phi_S$, $\Phi_{S^{'}}$ and $\Phi_{S^{''}}$, which can be candidates for the dark matter, should be sterile scalars interacting very weakly, or not interacting at all, with the SM fields including $\Phi_h$. That means that the interaction \eqref{Vphi3s} should be much smaller than the interaction \eqref{VPhi}. After an $A_4$ symmetry breaking, 
		$\Phi_h$ is seen as a single SM Higgs scalar coupled to three fermion generations with different coupling coefficients leading to different masses of the fermions.\\
	
	Imposing the extremum condition to $\Phi_h$,
	\begin{equation}
	\left. \frac{\partial V}{\partial \phi_{h_i}}\right 
	\vert _{\langle \phi_{h_i} \rangle=v_i}=0,~~~~ (i=1,2,3),
	\label{phi-min}
	\end{equation}
	we get an equation system of their VEV's
\begin{widetext}
	\begin{equation}
	\begin{cases}
	\mu_0^2+2(\lambda_1+\lambda_2) v_1^2 +(2 \lambda_1-\lambda_2+2 \lambda_4+\lambda_5) v_2^2+(2 \lambda_1-\lambda_2+2 \lambda_3+\lambda_5) v_3^2 =0, \\
	\mu_0^2+2(\lambda_1+\lambda_2) v_2^2+(2 \lambda_1-\lambda_2+2 \lambda_3+\lambda_5) v_1^2+(2 \lambda_1-\lambda_2+2 \lambda_4+\lambda_5) v_3^2 =0, \\
	\mu_0^2+2(\lambda_1+\lambda_2) v_3^2+(2 \lambda_1-\lambda_2+2 \lambda_3+\lambda_5) v_2^2+(2 \lambda_1-\lambda_2+2 \lambda_4+\lambda_5) v_1^2 =0.
	\end{cases}
	\label{hptv} 
	\end{equation}
\end{widetext}
	The latter equation system for general coefficients (general $\lambda_0$ and $\lambda_i$'s), 
	is equivalent to the constraints 
	\begin{equation}
	v_1^2=v_2^2=v_3^2:=v^2,
	\label{vev}
	\end{equation}
	leading to eight different choices, 
	$$\pm v_1=\pm v_2=\pm v_3=\pm v,$$ 
	of the VEV's structure of $\Phi_h$, where
	\begin{equation}
	v^2 = \frac{- \mu_0^2}{2(3 \lambda_1+\lambda_3+ \lambda_4+\lambda_5)}~\cdot
	\label{v2}
	\end{equation}
Due to the couplings of $\Phi_h$ to the SM fields, in particular, the SM gauge bosons, the VEV $v$ can have a value at the electroweak scale as in the SM. \\

In general, the Higgs fields defined above are flavour states but not mass-eigen states. To get their mass-eigen states one must diagonal the Higgs mass matrix appearing squared in the mass term 
	\begin{equation}
	V_M (\Phi) \sim {1\over 2}\left(
	\begin{array}{ccc}
	h_1 & h_2 & h_3
	\end{array}
	\right) M_h^2
	\left(
	\begin{array}{c}
	h_1 \\
	h_2 \\
	h_3
	\end{array}
	\right),
	\end{equation}
	where
	\begin{equation}
	\label{mHigg}
	M_h^2 = \left(
	\begin{array}{ccc}
	p & q & q \\
	q & p & q\\
	q & q & p
	\end{array}
	\right),
	\end{equation}
	with
	\begin{align*}
	p =& 2v^2 (\lambda_1+\lambda_2+\lambda_3+\lambda_4),  \\ 
	q =&   v^2 (2  \lambda_1-\lambda_2+\lambda_5).
	\end{align*}
	As \eqref{mHigg} is a real symmetric matrix, its eigenvalues, therefore, the scalar masses, 
	are always real. It is not difficult to see that the matrix 
	\begin{equation}
	U_H = \left(
	\begin{array}{ccc}
	\sqrt{1\over 3}  & -\sqrt{2\over 3} & 0 \\[2mm]
	\sqrt{1\over 3}  & \sqrt{1\over 6} & -\sqrt{1\over 2}  \\[2mm]
	\sqrt{1\over 3}  & \sqrt{1\over 6}  & \sqrt{1\over 2}
	\end{array}
	\right).
	\end{equation}
	can rotate the (squared) mass matrix $M_h^2$ to the diagonal matrix 
	\begin{equation}
	\label{mh}
	M_H^2 = \left(
	\begin{array}{ccc}
	m^2_{H1} & 0 & 0 \\[4mm]
	0 & m^2_{H2} & 0 \\[4mm]
	0 & 0 & m^2_{H3}
	\end{array}
	\right),
	\end{equation}
	where 
	\begin{align}
	m_{H_1}^2=p+2q=&2(3\lambda_1+\lambda_3+\lambda_4+\lambda_5)v^2=-\mu_0^2,
	\label{mH1}\\
	m_{H_2}^2=m_{H_3}^2=&~p-q=(3\lambda_2+2\lambda_3+2\lambda_4-\lambda_5)v^2  \nonumber \\
	=& -\mu^2_0-3(2\lambda_1-\lambda_2+\lambda_5)v^2,
	\label{mH23}
	\end{align}
	are masses of the Higgs mass-eigen states $H_i$ related to the flavour states $h_i$ 
	via the rotation 
	\begin{equation}
	\left(
	\begin{array}{c}
	h_1 \\
	h_2  \\
	h_3
	\end{array}
	\right)= U_H^t 
	\left(
	\begin{array}{c}
	H_1  \\
	H_2  \\
	H_3
	\end{array}
	\right).
	\end{equation}
A similar analysis can be done for other scalar fields. A more  detailed analysis on the scalar sector is very interesting but such as analysis, however, goes beyond the scope of this paper, and therefore, it will be a subject of a separate work.
%
%
\subsection{Lepton sector}
\hspace{0.5cm}
The lepton sector contains charged leptons and neutrinos. In the present model, 
the charged lepton masses can be generated by the Yukawa terms of the Lagrangian 
\begin{align}
\label{L1}
- \mathcal{L}_{Y_{cl}} = & ~ y_1 (\overline{\ell}_L \Phi_{h}) \ell_{R1} 
+ y_2(\overline{\ell}_L \Phi_{h})^{''} 
\ell_{R2} +y_3(\overline{\ell}_L \Phi_{h})^{'} \ell_{R3} +h.c. 
\end{align}
After an $A_4$ and gauge symmetry breaking, the above-given Yukawa terms become 
\begin{align}
\label{L2}
- \mathcal{L}_{Y_{cl}} = & y_1 ( v_1\overline{\ell}_{L1}+ v_2\overline{\ell}_{L2} 
+ v_3\overline{\ell}_{L3}) \ell_{R1}   \nonumber\\
&+ y_2 ( v_1\overline{\ell}_{L1}+\omega  v_2 \overline{\ell}_{L2} 
 +\omega^2  v_3\overline{\ell}_{L3}) 
\ell_{R2}   \nonumber\\
& + y_3(\overline{ v_1\ell}_{L1}+\omega^2  v_2\overline{\ell}_{L2} 
+\omega  v_3 \overline{\ell}_{L3}) 
\ell_{R3} + h.c., 
\end{align}
and  the charged leptons gain masses with 
the mass matrix 
\begin{equation}
M_{lept}=
\left(
\begin{array}{ccc}
y_1 v_1 & y_2 v_1 &  y_3 v_1 \\
y_1 v_2 & \omega y_2 v_2 &  \omega^2 y_3 v_2 \\
y_1 v_3 &\omega^2  y_2 v_3 &  \omega y_3 v_3
\end{array}
\right).
\end{equation}
Taking into account \eqref{vev}, we can, without loss of generality, 
assume \footnote{Another choice of the Higgs VEV satisfying \eqref{vev} 
gives a similar result.} that 
$\langle \Phi_h \rangle ^T=\frac{1}{\sqrt{2}} 
(v,v,v)$. The charged lepton mass matrix, then, takes the form
\begin{equation}
M_{lept}=\frac{1}{\sqrt{2}} U_L
\left(
\begin{array}{ccc}
m_e & 0 & 0 \\
0 & m_{\mu}  & 0 \\
0 & 0 & m_{\tau}
\end{array}
\right),
\end{equation}
where 
\begin{equation}
m_e=y_1 v, ~~ m_\mu = y_2 v, ~~m_\tau =y_3 v,
\end{equation}
are the charged lepton masses, and 
\begin{equation}
U_L= \frac{1}{\sqrt{3}}
\left(
\begin{array}{ccc}
1 &  1 &  1 \\
1 & \omega  &  \omega^2  \\
1 &\omega^2  &  \omega
\end{array}
\right).
\end{equation}
\vspace{2mm}

For the neutrinos, their masses can be generated by Yukawa terms of Dirac and Majorana  type. 
First, we deal with the Dirac Yukawa terms
%
	\begin{align}
	\label{NYukawaDirac}
	- \mathcal{L}_{Y_{\nu}}^D = &~ y^{\nu}_{Ta}\left(\bar{\ell_L}
	\widetilde{\Phi}_h \right)_{3_a}\cdot 
	N_T+y^{\nu}_{Tb}\left(\bar{\ell_L}\widetilde{\Phi}_h \right)_{3_s}\cdot N_T  \\ \nonumber
	& +y^{\nu}_{S}\left(\bar{\ell_L}  \widetilde{\Phi}_h \right)_{1}\cdot N_S  
	+y^{\nu}_{S'}\left(\bar{\ell_L} \widetilde{\Phi}_h \right)_{1''}\cdot N_{S'}   \\ \nonumber
	& +y^{\nu}_{S''}\left(\bar{\ell_L} 
	\widetilde{\Phi}_h \right)_{1'}\cdot N_{S''} +h.c..
	\end{align}
The Dirac mass matrix extracted from these Yukawa terms is 	
	\begin{equation}
	\label{M1}
	M_D =
	\left(
	\begin{array}{ccccccccc}
	0 & {y_{Ta}^\nu}v_3  & {y_{Tb}^\nu}v_2 & {y_S^\nu}v_1 & {y_{S'}^\nu}v_1
	& {y_{S''}^\nu}v_1 \\
	{y_{Tb}^\nu}v_3 &  0 &  {y_{Ta}^\nu}v_1  & {y_S^\nu}v_2  &
	{y_{S'}^\nu}v_2\omega^2  & {y_{S''}^\nu}v_2\omega  \\
	{y_{Ta}^\nu}v_2  & {y_{Tb}^\nu} v_1 & 0 & {y_S^\nu} v_3  &
	{y_{S'}^\nu}v_3\omega  &  {y_{S''}^\nu}v_3\omega^2
	\end{array}
	\right).
	\end{equation}
The Majorana Yukawa terms have the form 
\begin{widetext}
	\begin{align}
	\label{NYukawaMajo}
	- \mathcal{L}_{Y_{\nu}}^M =& ~y_{T_1}^M \left( \overline{N}_T N_T  \right)_{1}\Phi_S+y_{T_2}^M \left( \overline{N}_T N_T  \right)_{1^{'}}\Phi_{S^{''}}+y_{T_2}^M \left( \overline{N}_T N_T  \right)_{1^{''}}\Phi_{S^{'}}  \nonumber  \\
	&+y_1^M \left( \overline{N}_S N_S  \right)_{1} \Phi_S+y_2^M \left( \overline{N}_{S^{'}} N_{S^{''}} \right)_{1} \Phi_S   \nonumber  \\
	&+y_3^M \left( \overline{N}_{S^{'}} N_{S^{'}} \right)_{1^{''}} \Phi_{S^{'}}+y_4^M \left( \overline{N}_S N_{S^{''}} \right)_{1^{''}} \Phi_{S^{'}}  \nonumber  \\
	&+y_5^M \left( \overline{N}_{S^{''}} N_{S^{''}} \right)_{1^{'}} \Phi_{S^{''}}+y_6^M \left( \overline{N}_S N_{S^{'}} \right)_{1^{'}} \Phi_{S^{''}}+h.c..
	\end{align}
\end{widetext}
	The latter Lagrangian \eqref{NYukawaMajo} can be re-witten as 
	\begin{equation}
	\label{NYukawaMajo1}
	- \mathcal{L}_{Y_{\nu}}^M = \frac{1}{2} \overline{(N_{R})^c} \cdot M_{R} \cdot N_{R} +h.c.,
	\end{equation}
	where 
\begin{align}
N_T =& \left( N_{T1}, N_{T2}, N_{T3}\right)^T,   \\ 
N_{R} = & \left( N_{T1}, N_{T2}, N_{T3}, N_{S}, N_{S'}, N_{S''}   \right)^T,
\end{align}
	and 
	\begin{equation}
	\label{MMaj}
	M_R =
	\left(
	\begin{array}{ccccccccc}
	M_{11} & 0 & 0 & 0 & 0 & 0  \\
	0  & M_{22} & 0 & 0 & 0 & 0   \\
	0 & 0 & M_{33} & 0 & 0 & 0   \\
	0 & 0 & 0 & M_{44} & M_{45} & M_{46}   \\
	0 & 0 & 0 & M_{54} & M_{55} & M_{56}   \\
	0 & 0 & 0 & M_{64} & M_{65} & M_{66}   \\
	\end{array}
	\right)
	\end{equation}
is the Majorana mass matrix in the present case  	
with 
	\begin{align}
	M_{11}&=y_{T_1}^M\sigma_1+~~~y_{T_2}^M\sigma_2+~~~y_{T_3}^M\sigma_3,   \nonumber  \\
	M_{22}&=y_{T_1}^M\sigma_1+\omega^2 y_{T_2}^M\sigma_2+\omega ~ y_{T_3}^M\sigma_3   \nonumber,  \\
	M_{33}&=y_{T_1}^M\sigma_1+\omega ~ y_{T_2}^M\sigma_2+\omega^2 y_{T_3}^M\sigma_3   \nonumber,  \\
	M_{44}&=y_1^M\sigma_1,~~~M_{56}=M_{65}=y_2^M\sigma_1,    \nonumber   \\
	M_{55}&=y_3^M\sigma_2,~~~M_{46}=M_{64}=y_4^M\sigma_2,    \nonumber   \\
	M_{66}&=y_5^M\sigma_3,~~~M_{45}=M_{54}=y_6^M\sigma_3.
	\end{align}
It is not easy to diagonalize the Majorana mass matrix \eqref{MMaj} in its general form which, however, is not always physical. Therefore, we could make some assumptions which are physically reasonable and can simlify the diagonalization of the mass matrix \eqref{MMaj}.\\ 
	
We assume that interactions between the sterile neutrinos and each of the sterile scalars have strengths of the same order, that is,  
	\begin{itemize}
		\item
		$y_1^M=y_2^M$, that means $M_{44}=M_{56}=M_{65}$, 
		\item
		$y_3^M=y_4^M$, that means $M_{55}=M_{46}=M_{64}$, 
		\item
		$y_5^M=y_6^M$, that means $M_{66}=M_{45}=M_{54}$. 
	\end{itemize}
Thus the Majorana mass takes now the form 
	\begin{equation}
	\label{MMaj-2}
	M_R =
	\left(
	\begin{array}{ccccccccc}
	M_{11} & 0 & 0 & 0 & 0 & 0  \\
	0  & M_{22} & 0 & 0 & 0 & 0   \\
	0 & 0 & M_{33} & 0 & 0 & 0   \\
	0 & 0 & 0 & M_{44} & M_{66} & M_{55}   \\
	0 & 0 & 0 & M_{66} & M_{55} & M_{44}   \\
	0 & 0 & 0 & M_{55} & M_{44} & M_{66}   \\
	\end{array}
	\right).
	\end{equation}
	
Taking into account \eqref{NYukawaDirac} and \eqref{NYukawaMajo} we write the Yukawa terms for neutrinos 	
	\begin{align}
	\mathcal{L}_{Y_{\nu}}=\mathcal{L}_{Y_{\nu}}^D+ \mathcal{L}_{Y_{\nu}}^M
	\end{align}
as follows  
	\begin{equation}
	\mathcal{L}_{Y_{\nu}}= \frac{1}{2} \overline{n}_{L} 
	M_{\rm seesaw} (n_{L})^c+h.c,
	\end{equation}
where 
	\begin{equation}
	n_L=\left(\nu_L,(N_R)^c\right)^T,
	\end{equation}
and 
	\begin{equation}
	M_{\rm seesaw} =
	\left(
	\begin{array}{cc}
	0  & M_D  \\
	M_D^T & M_R
	\end{array}
	\right).
	\end{equation}
From here, we get the type-I see-saw mass matrix 	
	\begin{equation}
	\label{Seesawmass}
	M_{\nu}=-M_D^T (M_R)^{-1} M_D,
	\end{equation}
which in the charged-lepton diagonalizing basis becomes 
	\begin{equation}
	\mathcal{M}_{\nu}=U_L^{\dagger} M_{\nu} U_L^{\ast}=
	\left(
	\begin{array}{ccc}
	\label{masselement2}
	A & B  & C  \\
	B & E & D  \\
	C & D & F
	\end{array}
	\right),
	\end{equation}
Before finishing this section let us discuss the mass scales, $M_R$, $M_D$ and $M_\nu$, 
involved. The general ranges of these mass scales are given in Tab. \ref{MR}. According 
to the current experimental data \cite{Olive:2016xmw} the upper bound of the light 
neutrino masses (the scale of $M_\nu$) is below $10^{-1}$ eV = $10^{-10}$ GeV. Thus,  
if the scale of $M_D$ is known, we can estimate the mass scale $M_R$ in relation to the 
coupling coefficients in \eqref{NYukawaDirac} and \eqref{NYukawaMajo}, and vice versa, knowing $M_R$ we can determine 
$M_D$. While the scale $M_\nu$ is bounded today narrowly between 0 eV and $10^{-1}$ eV 
(even less), the scales of $M_R$ and  $M_D$ still vary in wider ranges covering the 
electroweak (EW) 
scale $10^2$ GeV as well as the LHC discovering potential at about 1 TeV = $10^3$ GeV. 
On searching for heavy Majorana neutrinos within the type-I see-saw mechanism with 
the ATLAS detector in $\sqrt{s}=8$ TeV $pp$ collisions \cite{Aad:2015xaa} 
the range of $M_R$ has been set to be between 100 -- 500 GeV, while the ranges given 
by CMS are 90 GeV $<M_R<200$ GeV (for $ee$ final states in $\sqrt{s}=7$ TeV $pp$ 
collisions) \cite{Chatrchyan:2012fla} and 40 GeV $<M_R<500$ GeV (for $\mu\mu$ final 
states in $\sqrt{s}=8$ TeV $pp$ collisions) \cite{Khachatryan:2015gha}. 
An $M_R$ at the GUT scale and an $M_D$ at the EW scale are often 
taken in a see-saw model (see, for example, \cite{Dinh:2006ia} and references therein) 
but, in general, the range of $M_R$, as mentioned earlier, could spread from the eV 
scale to the GUT scale (see, for example, \cite{Adhikari:2016bei} and references therein) 
and $M_D$, at least here, is not constrained yet by any other constraint besides 
that related to Yukawa coupling coefficients in \eqref{NYukawaDirac} and \eqref{NYukawaMajo}. A determination or 
estimation of the latter (via measurements of some processes such as those involved 
charged leptons) can shed a light on the scale of $M_D$ and thus on that of $M_R$. 
The branching ratios of a charged lepton decay tells us that 
these Yukawa coupling coefficients could be at least four magnitudes (10$^{-4}$) below 
that of an EW one, hence $M_D$ could be at the order 10$^{-2}$ GeV at most, thus, $M_R$ 
could be at the 10 TeV scale at most. Interestingly, the recent observation of 3.5 keV 
X-ray signals from several galaxies and galaxy clusters 
\cite{Bulbul:2014sua,Boyarsky:2014jta} could be explained by a decay of keV sterile 
neutrinos. The latter, if in a see-saw model, lead to very small (unless the VEV of the 
related scalar field(s) is small or the scale of $M_\nu$ is big enough) Dirac mass scale 
and coupling coefficients (see Tab. \ref{MR}), the measurement of which is still very 
difficult. \\
\begin{widetext}
\begin{center}
\begin{figure}[H] 
\begin{center}
\begin{tabular}{|c||c|c|c|c|c|c|c|c|c|}
\hline
 \textbf{Mass scale}              & $\mathbf{M_R}$ (GeV)       & $\mathbf{M_{\nu}}$ (GeV)   
 & $\mathbf{M_D=(M_{\nu} M_R )^{1/2}}$ (GeV) 
 &  $\mathbf{y_D = M_D v^{-1}}$    \\  \hline
GUT    &$ \sim 10^{15} $    &  $ < 10^{-10}$   &   $ \sim 10^2 \div 10^3$                           
        &     $ \sim 10^0 \div 10^1$         \\  \hline
 TeV    & $ \sim 10^{3} $      &  $ <  10^{-10}$    &     $ \sim 10^{-4} \div 10^{-3}$  
  &     $\sim  10^{-6} \div 10^{-5}$             \\  \hline
 GeV   & $ \sim 10^0 $      &  $ <  10^{-10}$    &    $ \sim 10^{-5}$                                
      &     $\sim  10^{-7}$             \\  \hline
 MeV   & $ \sim 10^{-3} $      &  $ <  10^{-10}$    &    $ \sim 10^{-7} \div 10^{-6}$ 
  &     $\sim  10^{-9} \div 10^{-8}$             \\  \hline
 keV    & $ \sim 10^{-6} $      &  $ <  10^{-10}$    &    $ \sim 10^{-8}$                        
    &    $\sim  10^{-10}$             \\  \hline
 eV    & $ \sim 10^{-9} $      &  $ <  10^{-10}$    &    $ \sim 10^{-10} \div 10^{-9}$                 
           &    $\sim  10^{-12} \div 10^{-11}$             \\  \hline
\end{tabular}
\caption{\small \label{MR} Mass scales of the model.}
\end{center}
\end{figure}
\end{center}
\end{widetext}

\section{Neutrino masses and mixing}
It is observed from \eqref{NYukawaDirac}, \eqref{NYukawaMajo1} and  \eqref{masselement2} that if 
\begin{equation}
y_S^{\nu}=y_{S^{'}}^{\nu}=y_{S^{''}}^{\nu},~~~y_{Ta}, y_{Tb} \ll y_S^{\nu}, ~~~M_{55}=M_{66}, 
\label{cTBM} 
\end{equation}	
the neutrino mass matrix having the form 
\begin{widetext}
\begin{equation}
\mathcal{M}_{\nu0}=\frac{1}{\Lambda}
\left(
\begin{array}{ccc}
3 (M_{44}+M_{66})(y_S^{\nu})^2  & -3 M_{66}(y_S^{\nu})^2  & -3 M_{66}(y_S^{\nu})^2  \\
-3 M_{66}(y_S^{\nu})^2  & -3 M_{66}(y_S^{\nu})^2 & 3 (M_{44}+M_{66})(y_S^{\nu})^2  \\
-3 M_{66}(y_S^{\nu})^2  & 3 (M_{44}+M_{66})(y_S^{\nu})^2 & -3 M_{66}(y_S^{\nu})^2
\end{array}
\right),
\label{Mnu0}
\end{equation}
\end{widetext}
where 	
	\begin{equation}
	\Lambda = (M_{44}-M_{66})(M_{44}+2 M_{66}),
	\end{equation}
can be diagonalized 
	\begin{equation}
	\diag(\mathcal{M}_{\nu 0}) = {(U'_{tbm})}^{T} \mathcal{M}_{\nu 0}U'_{tbm},
	\end{equation}
by the matrix 
	\begin{equation}
	U'_{tbm}=\left(
	\begin{array}{ccc}
	\sqrt{\frac{2}{3}} & \sqrt{\frac{1}{3}} & 0 \\
	-\sqrt{\frac{1}{6}} & \sqrt{\frac{1}{3}} & -\sqrt{\frac{1}{2}} \\
	-\sqrt{\frac{1}{6}} & \sqrt{\frac{1}{3}} & \sqrt{\frac{1}{2}}
	\end{array}
	\right).
\label{tbm}	
	\end{equation}
~\\
That means, the model with the condition \eqref{cTBM} is a TBM model.
Here, the matrix $U'_{tbm}$ can be converted to a more convenient form 
\begin{equation}
U_{tbm}=\left(
\begin{array}{ccc}
\sqrt{\frac{2}{3}} & \sqrt{\frac{1}{3}} & 0 \\
-\sqrt{\frac{1}{6}} & \sqrt{\frac{1}{3}} & \sqrt{\frac{1}{2}} \\
\sqrt{\frac{1}{6}} & -\sqrt{\frac{1}{3}} & \sqrt{\frac{1}{2}}
\end{array}
\right),
\end{equation}
However, as said above (and also in \cite{Ky:2016rzl,Dinh:2015tna}), the experimental PNMS mixing 
matrix has a small deviation from a TBM form, therefore, 
the former can be presented as a perturbation around the latter. It follows from the fact 
that the difference, say $\Delta U$, between the experimental PNMS mixing matrix 
\cite{Olive:2016xmw} and the TBM one is just a small correction to the latter 
(upto a phase factor): 
\begin{equation}
\Delta U=
\left(
\begin{array}{ccc}
 0.006 & -0.029 & 0.153 e^{-i \delta} \\
 -0.008-0.084 e^{i \delta} & 0.047\, -0.056 e^{i \delta} & -0.054 \\
 -0.041-0.095 e^{i \delta} & 0.027\, -0.064 e^{i \delta} & 0.034 \\
\end{array}
\right).
\end{equation}
Therefore, the (actual) neutrino mass matrix diagonalisable by the (experimental) 
PMNS mixing matrix, can be developed pertubatively around the TBM one 
\cite{Ky:2016rzl,Dinh:2015tna, Brahmachari:2014npa}.
As the present model under the condition \eqref{cTBM} becomes a TBM model, a realistic 
model can be obtained by imposing a condition slightly differing from  \eqref{cTBM}. In other 
words, to obtain a realistic model we could replace \eqref{cTBM} by another condition 
which at the first order of approximation reads
%
	\begin{align}
	& y_{S^{'}}^{\nu} = y_{S}^{\nu}+\epsilon_1,~~~ y_{S^{''}}^{\nu}= y_{S}^{\nu}+\epsilon_2,
~~~	M_{55} = M_{66}+\sigma,
	\end{align}
with $\epsilon_i$, $i=1,2$, and $\sigma$ are small number (i.e., $|\epsilon_i|, |\sigma| \ll 1$).	
From here we can develope perturbation of the Dirac and the Majorana mass matrix $M_D$ and $M_R$ around their TBM limit $M_{D_0}$ and $M_{R_0}$ respectively, 
	\begin{equation}
	M_D = M_{D_0}+\Delta_D
	\end{equation}
and  
	\begin{equation}
	M_R^{-1}=M_{R_0}^{-1}+\Delta_R, 
	\end{equation} 
where  
$\Delta_D$ and $\Delta_R$ are small corrections to $M_{D_0}$ and $M_{R_0}^{-1}$, respectively.	
 Now the neutrino mass matrix has the following (perturbative) form 
	\begin{equation}
	\mathcal{M}_{\nu} = (M_{D_0}+\Delta_D)^T(M_{R_0}^{-1}+\Delta_R)(M_{D_0}+\Delta_D)
	\end{equation}
which can be re-written as 	
	\begin{equation}
	\mathcal{M}_{\nu} = \mathcal{M}_{\nu 0} + \mathcal{W},
	\label{Mtach}
	\end{equation}
where 
	\begin{equation}
	\mathcal{M}_{\nu 0} = M_{D_0}^TM_{R_0}^{-1}M_{D_0},
	\label{Mtach0}
	\end{equation}
is the TBM mass matrix \eqref{Mnu0}	
and ${\cal W}$ is a deviation from the latter 
	\begin{equation}
	{\cal W} =\frac{1}{\Lambda^{'}} \left(
	\begin{array}{ccc}
	e_1 & e_2 & e_2^{\ast}  \\
	e_2 & e_4 & e_3  \\
	e_2^{\ast} & e_3 &  e_4^{\ast}
	\end{array}
	\right),
	\end{equation}
with 
\begin{widetext}
	\begin{equation}
	\Lambda^{'} = \frac{ 2(M_{44}-M_{66}) \left[ (M_{44}-M_{66})(M_{44}+2M_{66})-3\sigma M_{66} \right] }{y_S^{\nu}}
	\end{equation}
and 
	\begin{align}
	e_1 = &-\frac{4(\epsilon_1+\epsilon_2) (M_{44}-M_{66}) \left[ M_{66}(2M_{66}+\sigma)-(M_{44}+M_{66})(M_{44}-\sigma) \right]}{(M_{44}+2 M_{66})}  \nonumber  \\
	&+\frac{6y_{S}^{\nu}(M_{44}-M_{66})^2}{(M_{44}+2 M_{66})},  \nonumber  \\
	e_2 = & ~(\epsilon_1+\epsilon_2)(M_{66}-M_{44})(M_{66}+2M_{44})+3\sqrt{3} i(\epsilon_1-\epsilon_2)M_{66}(M_{44}-M_{66})  
	\nonumber  \\
	&+2\sigma(\epsilon_1-2\epsilon_2) (\omega M_{44}-M_{66})-2\sqrt{3} i \sigma (2\epsilon_1-\epsilon_2)M_{66},  \nonumber  \\
	e_3=&-2(\epsilon_1+\epsilon_2)(M_{44}-M_{66})(M_{44}+2M_{66})+2\sigma(\epsilon_1-2\epsilon_2)M_{44}-2\sigma(\epsilon_2-5\epsilon_1)M_{66},
	\nonumber  \\
	e_4= & ~2(M_{44}+2M_{66}) \left[ 2(\epsilon_1+\epsilon_2)(M_{44}-M_{66})- 3\omega \sigma y_{S}^{\nu} \right] \nonumber  \\
	& +4 \sigma (\epsilon_1+\epsilon_2)(2 \omega M_{66}-\omega^2 M_{44}-3 M_{66}).
	\end{align}
\end{widetext}

The TBM mixing matrix can be expressed in terms of eigen-vectors $|n^0\rangle$, 
$n=1,2,3$, of $\mathcal{M}_{\nu 0}$ and also $\mathcal{M}_{\nu 0}^{\dagger} \mathcal{M}_{\nu 0}$ as follows 
\begin{equation}
U_{TBM} =
\left(|1^0\rangle,|2^0\rangle,|3^0\rangle \right).
\end{equation}
The eigen-vectors $|n\rangle$ of the matrix $\mathcal{M}_{\nu}^{\dagger} \mathcal{M}_{\nu}$ perturbatively 
developed around $|n^0\rangle$ have the form (taken upto the first perturbation order)
\begin{equation}
|n\rangle=|n^0\rangle+\sum_{k\neq n}\lambda_{nk}|k^0\rangle +...,
\end{equation}
where
\begin{align}
\lambda_{nk}=&(|m_n^0|^2-|m_k^0|^2)^{-1}V_{nk},  \\
V_{nk}=& \langle n^0|\mathcal{M}_{\nu 0}^{\dagger} \mathcal{W} + \mathcal{W}^{\dagger} \mathcal{M}_{\nu 0} |k^0\rangle,
\end{align}
and $|m_i^0|^2$, $i=1,2,3$, are eigenvalues of $\mathcal{M}_{\nu 0}^{\dagger} \mathcal{M}_{\nu 0}$.\\

Now we can diagonalize the neutrino mass matrix square $\mathcal{M}_{\nu}^{\dagger} \mathcal{M}_{\nu}$,  
\begin{equation}
\label{MUrealation}
\tilde {U}^\dagger \mathcal{M}_{\nu}^{\dagger} \mathcal{M}_{\nu} \tilde {U}
= \text{diag}~(|m_1|^2,|m_2|^2,|m_3|^2),
\end{equation}
by the matrix 
\begin{widetext}
\begin{equation}
\label{PMNSModel}
\tilde {U} =
\begin{matrix}
\left(
\begin{array}{llr}
\sqrt{\frac{2}{3}}+\sqrt{\frac{1}{3}}x^{*} & \sqrt{\frac{1}{3}}- \sqrt{\frac{2}{3}}x& 
-\sqrt{\frac{2}{3}}y-\sqrt{\frac{1}{3}}z \\
-\sqrt{\frac{1}{6}}+\sqrt{\frac{1}{3}}x^{*}+\sqrt{\frac{1}{2}}y^{*} & \sqrt{\frac{1}{3}}
+\sqrt{\frac{1}{6}}x+\sqrt{\frac{1}{2}}z^{*} & \sqrt{\frac{1}{2}}+\sqrt{\frac{1}{6}}y
-\sqrt{\frac{1}{3}}z\\
\sqrt{\frac{1}{6}}-\sqrt{\frac{1}{3}}x^{*}+\sqrt{\frac{1}{2}}y^{*} & -\sqrt{\frac{1}{3}}
-\sqrt{\frac{1}{6}}x+\sqrt{\frac{1}{2}}z^{*} & \sqrt{\frac{1}{2}}-\sqrt{\frac{1}{6}}y
+\sqrt{\frac{1}{3}}z
\end{array}
\right),
\end{matrix}
\end{equation}
\end{widetext}
with
\begin{equation}
x =\lambda_{12}, \hspace{0.2cm}
y =\lambda_{13}, \hspace{0.2cm}
z =\lambda_{23}.
\end{equation}
~\\
Any matrix $U'$, related to $\tilde {U}$ by the relation $U'=\tilde {U}\times P'$, 
where 
$P'={\rm diag}\left(e^{i\alpha'_1/2},e^{i\alpha'_2/2},e^{i\alpha'_3/2}\right),$ 
with arbitrary $\alpha'_1$, $\alpha'_2$, $\alpha'_3$, is also able to diagonalize 
$\mathcal{M}_{\nu}^{\dagger} \mathcal{M}_{\nu}$.
Therefore, a PMNS matrix $U_{PMNS}$ can be defined upto a phase matrix $P'$, 
\begin{equation}
\label{PMNSModel0}
U_{PMNS}=\tilde {U}\times P',
\end{equation}
Writing $P'$ as 
$P'=e^{i\alpha'_3/2}{\rm diag}\left(e^{i(\alpha'_1-\alpha'_3)/2},e^{i(\alpha'_2-\alpha_3)/2},1\right)
\equiv e^{i\alpha'_3/2}{\rm diag}\left(e^{i{\alpha'}_{21}/2},e^{i{\alpha'}_{31}/2},1\right)$,  
we see that the phase $\alpha'_3$ can be removed by redefining the overall field phase as done 
in the case of $P$. Thus, one can write $P'$ in a form similar to that of $P$ in \eqref{p}: 
\begin{equation}
P'={\rm diag}\left(e^{i{\alpha'}_{21}/2},e^{i{\alpha'}_{31}/2},1\right).
\end{equation}
It is important to mention that $x$, $y$ and $z$, in general, are complex numbers, 
so do the matrix elements of $\tilde{U}$, therefore, not only the mixing angles 
but also the Majorana phases can get a small correction from $\tilde{U}$. The 
latter, through complex $x$, $y$ and $z$, has six degrees of freedom (DOF's). Four 
of them parametrize the three mixing angles and the CPV phase. The other 
two DOF's, namely, two phases, 
can be used to synchronize the phases of $P'$ with those of $P$. Therefore, 
two of $x$, $y$ and $z$ (or two of their independent linear combinations) 
can be taken  to be real. Comparing \eqref{PMNS} with \eqref{PMNSModel}, 
we see that it is reasonable to choose a real $x$.\\

\noindent Using the trivial properties of the PMNS matrix given in \eqref{PMNS}, 
$\tan^2\theta_{12}=|U_{12}|^2/|U_{11}|^2$, 
$\sin\theta_{13}=|U_{13}|$,
$\tan^2\theta_{23}=|U_{23}|^2/|U_{33}|^2$, applied to \eqref{PMNSModel}, we obtain 
at the first perturbation order the following relations:
\begin{equation}
\label{relat01}
\tan^2\theta_{12}\approx \frac{1-2\sqrt{2}{\rm Re}(x)}{2+2\sqrt{2}{\rm Re}(x)},
\end{equation}
\begin{equation}
\sin\theta_{13}=\left|\sqrt{\frac{2}{3}}y+\sqrt{\frac{1}{3}}z\right|,
\end{equation}
\begin{equation}
\label{relat02}
\tan^2\theta_{23}\approx \frac{1+2\sqrt{2}{\rm Re}\left(\sqrt{\frac{1}{6}}y-
\sqrt{\frac{1}{3}}z\right)}{1-2\sqrt{2}{\rm Re}\left(\sqrt{\frac{1}{6}}y
-\sqrt{\frac{1}{3}}z\right)}.
\end{equation}
Further, from Eqs. (\ref{PMNS0}) and (\ref{PMNSModel0}), it is easy to get 
\bea
\label{realx}
\left(\sqrt{\frac{2}{3}}+\sqrt{\frac{1}{3}}x^{*}\right)e^{i\alpha'_{21}/2}
=c_{12}c_{13}e^{i\alpha_{21}/2},\\[3mm]
\left(\sqrt{\frac{1}{3}}- \sqrt{\frac{2}{3}}x\right)e^{i\alpha'_{31}/2}
=s_{12}c_{13}e^{i\alpha_{31}/2},\\[3mm]
\left(-\sqrt{\frac{2}{3}}y-\sqrt{\frac{1}{3}}z\right)
=s_{13}e^{-i\delta}.
\eea
In general, as $x$ is complex, $\alpha'_{21}\neq \alpha_{21}$ and 
$\alpha'_{31}\neq \alpha_{31}$, but in the case of a real $x$ we 
have (at the first order of perturbation) 
$\alpha'_{21}= \alpha_{21}$ and $\alpha'_{31}= \alpha_{31}$.\\

Below, to check how our model works we will consider the case of a real parameter 
$x$, that is $P'=(e^{i\alpha_{21}/2},e^{i\alpha_{31}/2},1)=P$, as argued above. 
Now, the PMNS matrix  (\ref{PMNSModel0}) gets the form 
\begin{widetext}
\begin{equation}
\label{PMNSModelrealx}
U_{PMNS} =
\begin{matrix}
\left(
\begin{array}{lll}
\sqrt{\frac{2}{3}}+\sqrt{\frac{1}{3}}x & \sqrt{\frac{1}{3}}- \sqrt{\frac{2}{3}}x& 
-\sqrt{\frac{2}{3}}y-\sqrt{\frac{1}{3}}z \\
-\sqrt{\frac{1}{6}}+\sqrt{\frac{1}{3}}x+\sqrt{\frac{1}{2}}y^{*} & \sqrt{\frac{1}{3}}
+\sqrt{\frac{1}{6}}x+\sqrt{\frac{1}{2}}z^{*} & \sqrt{\frac{1}{2}}+\sqrt{\frac{1}{6}}y
-\sqrt{\frac{1}{3}}z\\
\sqrt{\frac{1}{6}}-\sqrt{\frac{1}{3}}x+\sqrt{\frac{1}{2}}y^{*} & -\sqrt{\frac{1}{3}}
-\sqrt{\frac{1}{6}}x+\sqrt{\frac{1}{2}}z^{*} & \sqrt{\frac{1}{2}}-\sqrt{\frac{1}{6}}y
+\sqrt{\frac{1}{3}}z
\end{array}
\right)\times P.
\end{matrix}
\end{equation}
\end{widetext}
Deriving $x$, $y$ and $z$ by matching the matrix elements $(U_{PMNS})_{11}$, 
$(U_{PMNS})_{13}$ and $(U_{PMNS})_{23}$ in Eq. (\ref{PMNSModelrealx}) with the 
corresponding elements of the experimentally measured PMNS matrix at the current 
best fit value of neutrino oscillation data \cite{Olive:2016xmw,Capozzi:2017ipn, Esteban:2016qun}, 
then, inserting them back in Eq. (\ref{PMNSModelrealx}), we obtain the PMNS matrix 
given by the model 
\begin{widetext}
\begin{equation}
U_{PMNS}^{model} =
\begin{matrix}
\left(
\begin{array}{lll}
0.8221 & 0.5695& 0.1530\,e^{-i\delta} \\
 -0.4337-0.0883\,e^{i\delta}& 0.6252-0.0624\,e^{i\delta} & 0.6533\\
0.3716-0.0883\,e^{i\delta} & -0.5373-0.0624\,e^{i\delta} & 0.7609
\end{array}
\right)\times P.
\end{matrix}
\end{equation}
\end{widetext}
In the following, to check the present model we are going to analyze and discuss in 
the model scenario several physics quantities, which can be verified experimentally, 
such as the neutrinoless double beta decay effective mass $|\langle m_{ee}\rangle|$, 
CPV phase $\delta_{CP}\equiv \delta$ and Jarlskog parameter $J_{CP}$, using the PMNS 
matrix (\ref{PMNSModel}) matched with the current neutrino oscillation data 
\cite{Olive:2016xmw,Capozzi:2017ipn, Esteban:2016qun}. \\

Neutrinoless double beta decay ($0\nu \beta\beta$) is a process of emitting 
two electrons from relevant nuclei without producing any (anti)neutrino, 
thus violating the lepton number by 2 units 
\cite{Elliott:2004hr,Avignone:2007fu,Rodejohann:2011mu,Elliott:2002xe,
Bilenky:2002aw,Bilenky:2012qi,GomezCadenas:2011it,Schwingenheuer:2012jt}. 
The ($0\nu \beta\beta$) decay together with the radiative emission of neutrino 
pair (RENP) from atom \cite{Yoshimura:2011ri,Dinh:2012qb, Fukumi:2012rn} are 
the only two, so far, proposed processes, used for construction of experiments 
for determination of the nature of neutrinos (if they are of Dirac- or Majorana 
type). With its importance in understanding neutrino properties, the phenomenon 
of the ($0\nu \beta\beta$) decay has attracted interest of and has been studied 
by a large number of physicists in both theoretical and experimental aspects. 
Presently, the most sensitive experiments give an upper bound on 
the ($0\nu \beta\beta$) decay effective mass to be about 
$|\langle m_{ee}\rangle|<0.2-0.6$ eV (Heidelberg-Moscow 
\cite{KlapdorKleingrothaus:2000sn}), or $0.3-0.71$ eV (COURICINO \cite{Andreotti:2010vj}), 
or $0.14-0.38$ eV (EXO-200 \cite{Auger:2012ar}),
while the next generation of the experiments \cite{Bellini:2009zw,Gornea:2011zz} 
may obtain a signal of the decay if $|\langle m_{ee}\rangle|$ is not smaller than 
$(2-5)\times 10^{-2}$ eV. 
In the scenario of the present model, the ($0\nu \beta\beta$) decay effective mass 
has the form 
{\footnote{~See \cite{Bilenky:1987ty} for a detailed analysis on $\langle m_{ee}\rangle$ but here the smallness of the mixing matrix elements $U_{ek}$ and large values of masses $m_k$, for $k\ge 4$, are taken into account.}}
\begin{equation}
|\langle m_{ee}\rangle|=\left|m_1 U_{e1}^2+m_2 U_{e2}^2e^{i\alpha_{21}}
+m_3 U_{e3}^2e^{i\alpha_{31}}\right|.
\end{equation}
The dependence of $|\langle m_{ee}\rangle|$ on the PMNS mixing matrix 
and the lightest active neutrino mass $m_0$ is depicted in Fig. \ref{Meedata}.\\
\begin{widetext}
\begin{center}
\begin{figure}[H]
\begin{center}
\begin{tabular}{cc}
\includegraphics[width=6cm,height=6cm]{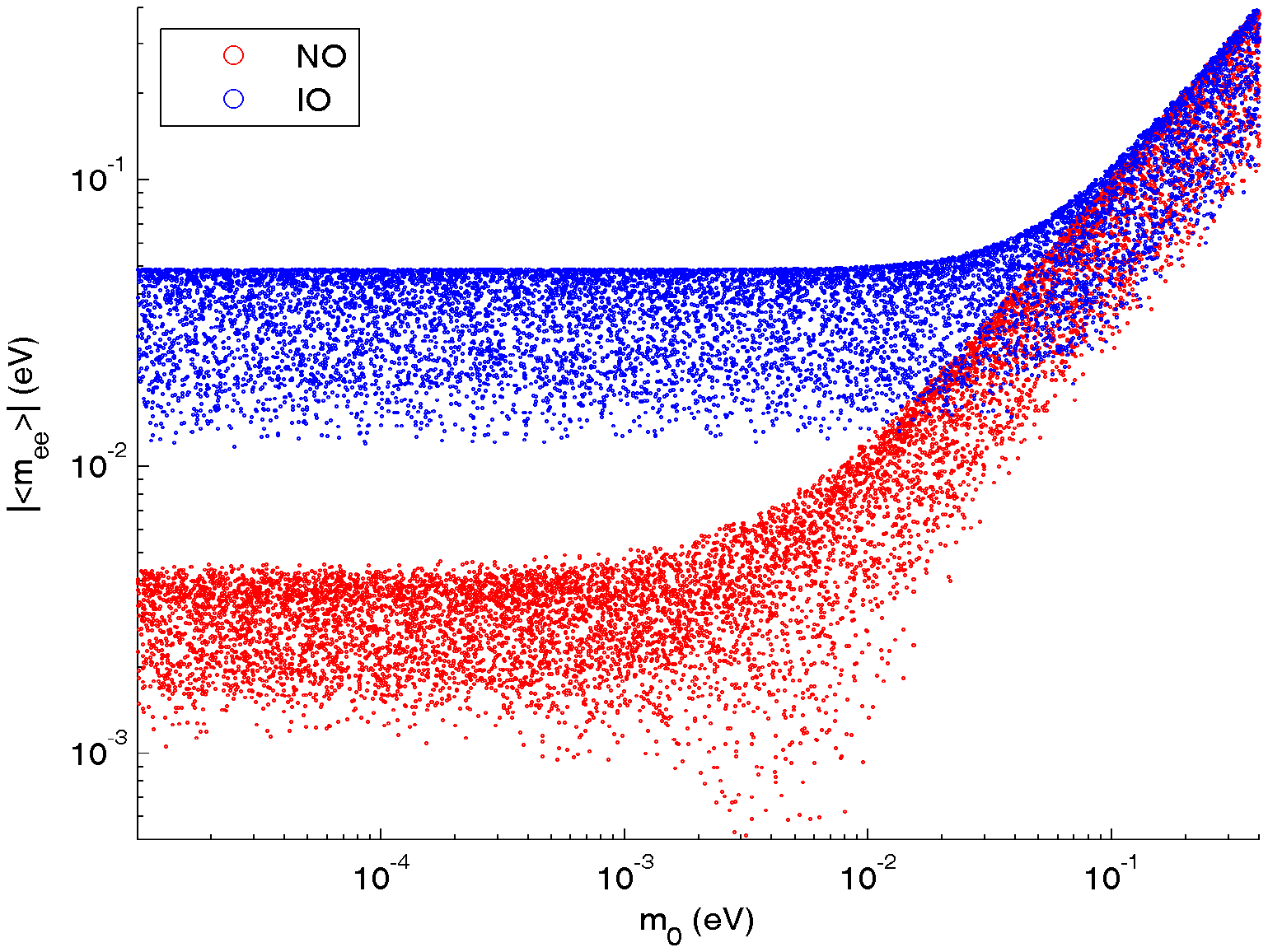}~~~&~~~
\includegraphics[width=5.7cm,height=5.7cm]{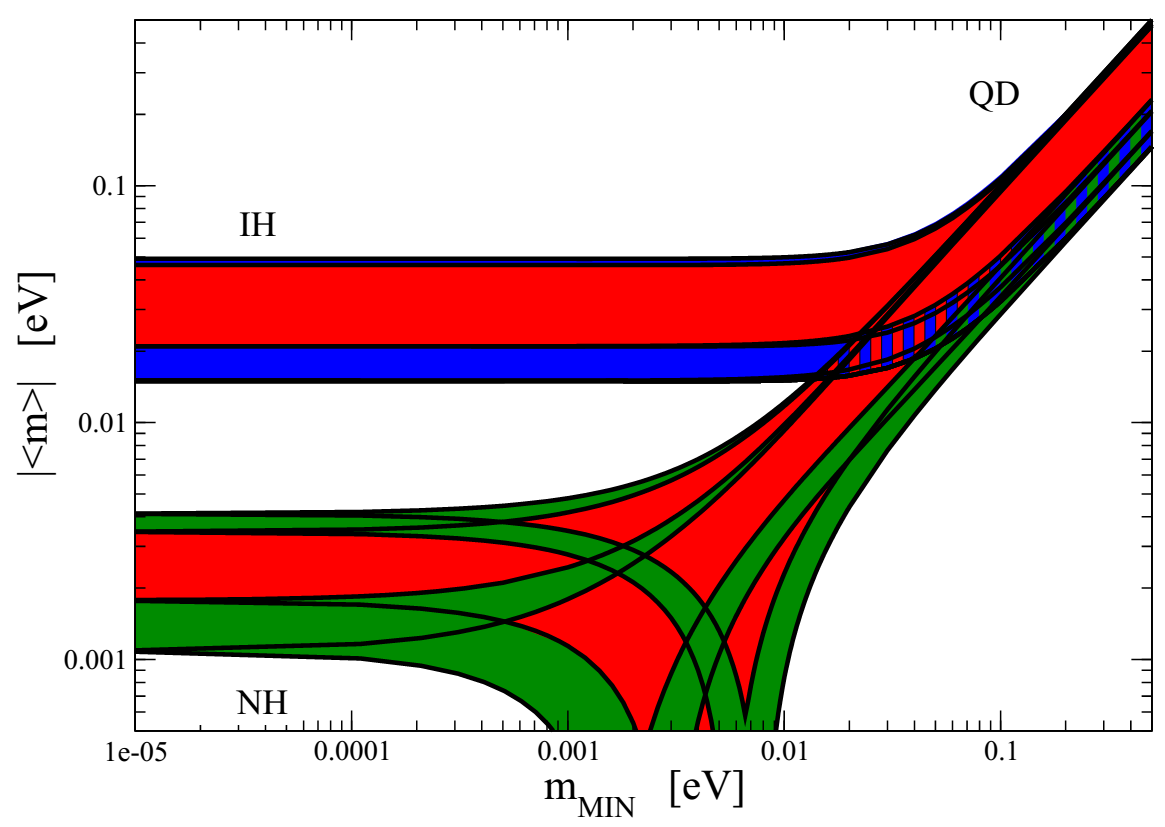}\\
\end{tabular}
\caption{The ($0\nu \beta\beta$) effective mass $|\langle m_{ee}\rangle|$ as a 
function of the lightest neutrino mass. \textit{Left panel}: Plot obtained by 
using (\ref{PMNSModelrealx}) with neutrino mixing angles taken arbitrary in 
$3\sigma$ ranges, while the phases $\delta$, $\alpha_{21}$ and $\alpha_{31}$ 
varying in $[0,2\pi]$. \textit{Right panel}: Plot taken from \cite{Olive:2016xmw}.}
\label{Meedata}
\end{center}
\end{figure}
\end{center}
\end{widetext}
To measure the Dirac CPV phase $\delta$ is a challenge but with the nonzero value of 
$\theta_{13}$ obtained by recent experiments, the measurement of $\delta$ becomes more 
realistic. The simplest and direct strategy, which has been arranged experimentally, 
is to determine the difference between the probabilities of a neutrino transition and 
an anti-neutrino transition 
\cite{Barger:1980jm, Pakvasa:1980bz},
\begin{align}
A_{CP}^{(\alpha,\beta)}= &P(\nu_\alpha\rightarrow\nu_\beta)
-P(\bar{\alpha}_\alpha\rightarrow\bar{\nu}_\beta)  \nonumber \\
=& -16J_{\alpha\beta}\sin\Delta_{12}\sin\Delta_{23}\sin\Delta_{31},
\end{align}
where $\Delta_{ij}\equiv \Delta m^2_{ij}L/4E$, and $J_{\alpha\beta}$ is the Jarlskog 
invariant quantity \cite{Jarlskog:1985ht},
\begin{equation}
J_{\alpha\beta}=J_{CP}={\rm Im}\{U_{\alpha 1}^*U_{\beta 2}^*U_{\alpha 1}U_{\beta 2}\}
=J_{CP}^{max}\sin\delta.
\label{J}
\end{equation}
Using the canonical parametrization of the PMNS matrix expressed in Eq. (\ref{PMNS}), 
it is not difficult to obtain
\begin{equation}
J_{CP}^{max}=\cos\theta_{12}\sin\theta_{12}\cos\theta_{23}\sin\theta_{23}
\cos^2\theta_{13}\sin\theta_{13}.
\label{Jmax}
\end{equation}
For the value of $\theta_{13}$ obtained by recent experiments, $J_{CP}^{max}$ takes 
value in the interval $[0.032-0.042]$ with neutrino oscillation angles varying in 
$3\sigma$ allowed ranges (learn more in Fig. 
\ref{JCPfigure}). We see that the most 
possible values of $\delta$ are located around $\pi$/2 (=90$^\circ$) and $3\pi$/2 
(=270$^\circ$), where $|J_{CP}|$ would take a value, which in fact is the maximal 
value $J_{CP}^{max}$, between $0.032 - 0.042$.\\
\begin{widetext}
\begin{center}
\begin{figure}[H]
\begin{center}
\begin{tabular}{cc}
\includegraphics[width=6.5cm,height=5.5cm]{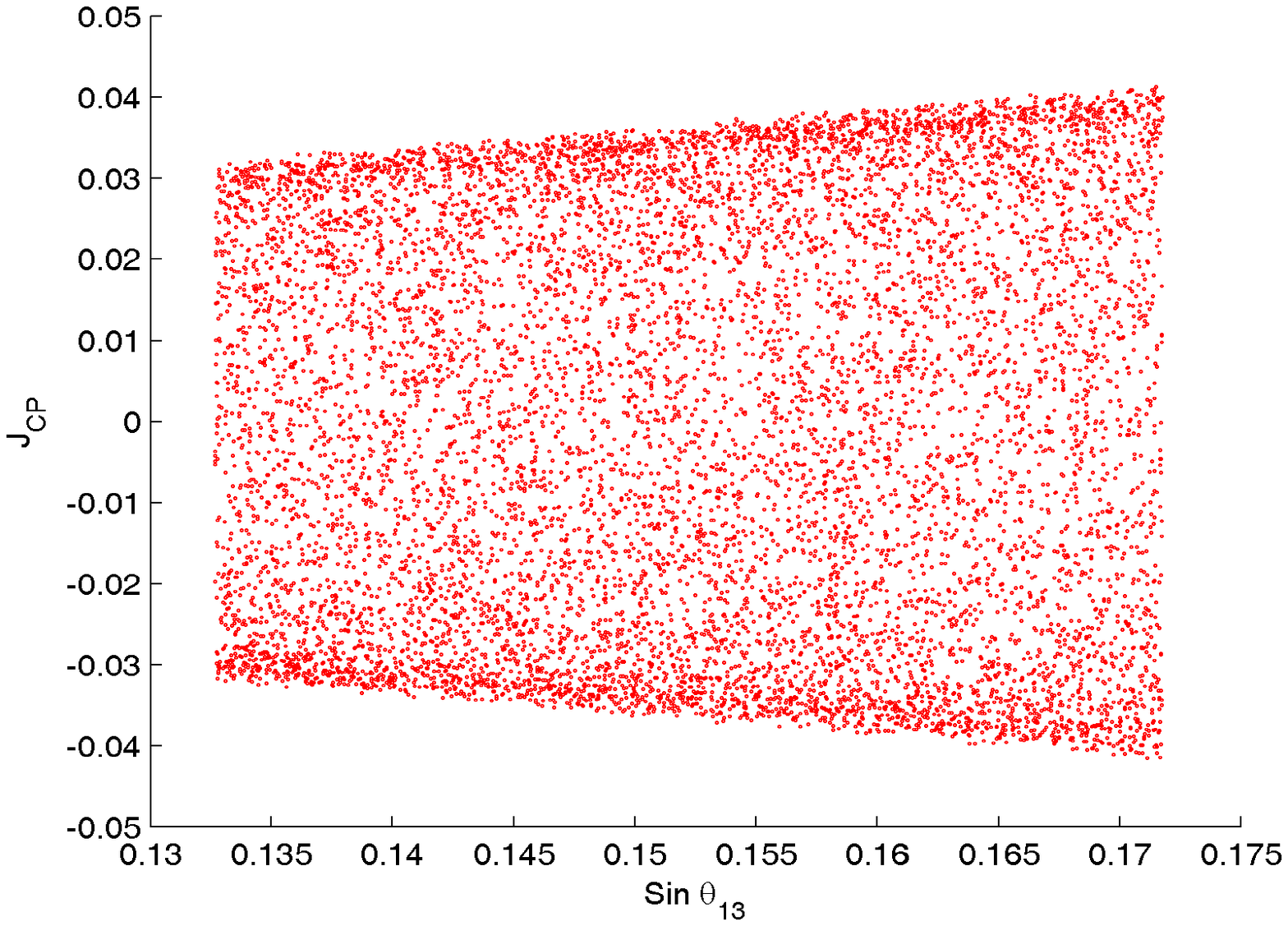}~~~&~~~
\includegraphics[width=6.5cm,height=5.5cm]{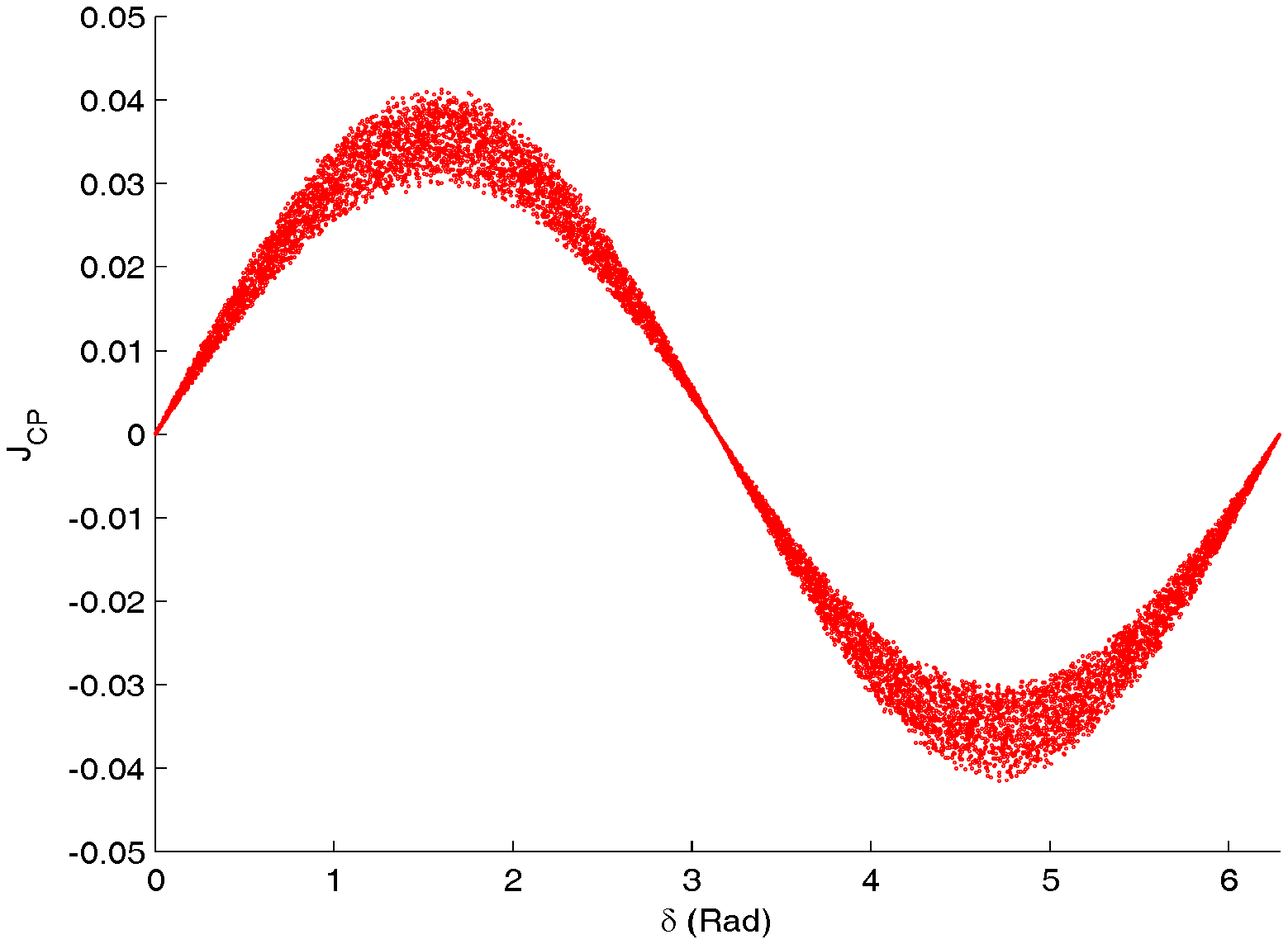} \\
\end{tabular}
\caption{\small $J_{CP}$ as a function of $\theta_{13}$ (\textit{left panel}) and  $\delta_{CP}$ 
(\textit{right panel}) with neutrino mixing angles taken in $3\sigma$ allowed ranges, and 
the CPV phase varying in $[0,2\pi]$.}
\label{JCPfigure}
\end{center}
\end{figure}
\end{center}
\end{widetext}
%
When $x$ is real, it is easy to see from Eqs. (\ref{relat01})--(\ref{relat02}), 
the value of the mixing angle $\theta_{12}$ depends only on $x$, while the other 
two mixing angles $\theta_{13}$ and $\theta_{23}$ as well as the Dirac CPV phase 
$\delta$ are determined by $y$ and $z$ via, for example, the equations  
\begin{align}
\label{relat03}
\sin\theta_{13}e^{-i\delta}=&-\sqrt{\frac{2}{3}}y-\sqrt{\frac{1}{3}}z,  \\
\tan^2\theta_{23}=& \frac{1+2\sqrt{2}{\rm Re}\left(\sqrt{\frac{1}{6}}y
-\sqrt{\frac{1}{3}}z\right)}{1-2\sqrt{2}{\rm Re}
\left(\sqrt{\frac{1}{6}}y-\sqrt{\frac{1}{3}}z\right)}.
\end{align}
Since $y$ and $z$ are complex numbers, they contain four parameters. One of these 
parameters can be used to synchronize a Majorana phase (as mentioned above), two 
are used to fix the mixing angles $\theta_{13}$ and $\theta_{23}$, and the 
remaining one is used for the Dirac CPV phase $\delta$. Therefore, $\delta$ can 
theoretically take any value in the range of $[0,2\pi]$.\\

Let us consider some specific values of parameters $x$, $y$, for example $y=0$ or $z=0$. 
In case $y=0$ and $z$ arbitrary (or $z=0$ and $y$ arbitrary), we have only two free 
parameters to fix 3 measurements $\theta_{13}$, $\theta_{23}$ and $\delta$, thus $\delta$ 
can be expressed in term of  $\theta_{13}$, $\theta_{23}$. For $y=0$, we have
\begin{align}
\label{relaty0}
-\sqrt{\frac{1}{3}}z=& \sin\theta_{13}e^{-i\delta},  \\
\tan^2\theta_{23}=& \frac{1+2\sqrt{2}{\rm Re}\left(-\sqrt{\frac{1}{3}}z\right)}{1-2\sqrt{2}{\rm Re}
\left(-\sqrt{\frac{1}{3}}z\right)}.
\end{align}
Eliminating $z$ in Eqs. (\ref{relaty0}), we obtain the constraint  
\begin{equation}
\sin\theta_{13}\cos\delta=\frac{1}{2\sqrt{2}}
\frac{\tan^2\theta_{23}-1}{\tan^2\theta_{23}+1}.
\label{sdelta}
\end{equation}
We can find $\delta$ from \eqref{sdelta} for a given set of the mixing 
angles (see their experimental data in Tab. \ref{Synopsis}). If a value $\delta_0$ is 
a solution of \eqref{sdelta} so is $2\pi-\delta_0$, therefore, it is enough to discuss 
one of them. This constraint gives, for example, at the best fit value (BFV) of the 
mixing angles $\theta_{13}$ and $\theta_{23}$  in an NO (and similarly for 
an IO), $\cos\delta=-0.291$, corresponding to the value $\delta\approx 4.414$ (and also 
$\delta\approx 1.866$). In general, this value of $\delta$ is not its mean value 
(BFV) but can give a rough estimation of the latter. \\

Based on the constraint 
\eqref{sdelta}, the distributions 
of $\delta$ for a normal ordering (NO) and an inverse ordering (IO), are depicted 
in Fig. \ref{Delta}, followed by Fig. \ref{Delta-s13}, where $\delta$ as function of 
$\theta_{13}$ is plotted also for both cases. 
For each of these distributions generated by 10000 events, the Dirac CPV 
phase $\delta$, and, thus, the Jarlskog parameter $J_{CP}$ (plotted in Figs. \ref{Jcp} 
and \ref{Jcp-s13}), is numerically calculated event by event with an input ($s_{ij}$) 
taken randomly on the base of a Gaussian distribution characterized by an experimental 
mean value (best fit value) and sigmas given in Tab. \ref{Synopsis}. 
These distributions $\delta$ have a mean values 
$\bar{\delta}_{NO}=4.417\approx 1.41\pi$ (for an NO) and 
$\bar{\delta}_{IO}=4.616\approx 1.47\pi$ (for an IO) 
which are close to the global fits at $\delta=\delta_{NO}\equiv 1.39\pi$ and at 
$\delta=\delta_{IO}\equiv 1.31\pi$, 
respectively \cite{Olive:2016xmw,Capozzi:2017ipn, Esteban:2016qun}. 
\begin{widetext}
\begin{center}
\begin{figure}[H]
\begin{center}
\begin{tabular}{cc}
\includegraphics[width=6.5cm,height=5.5cm]{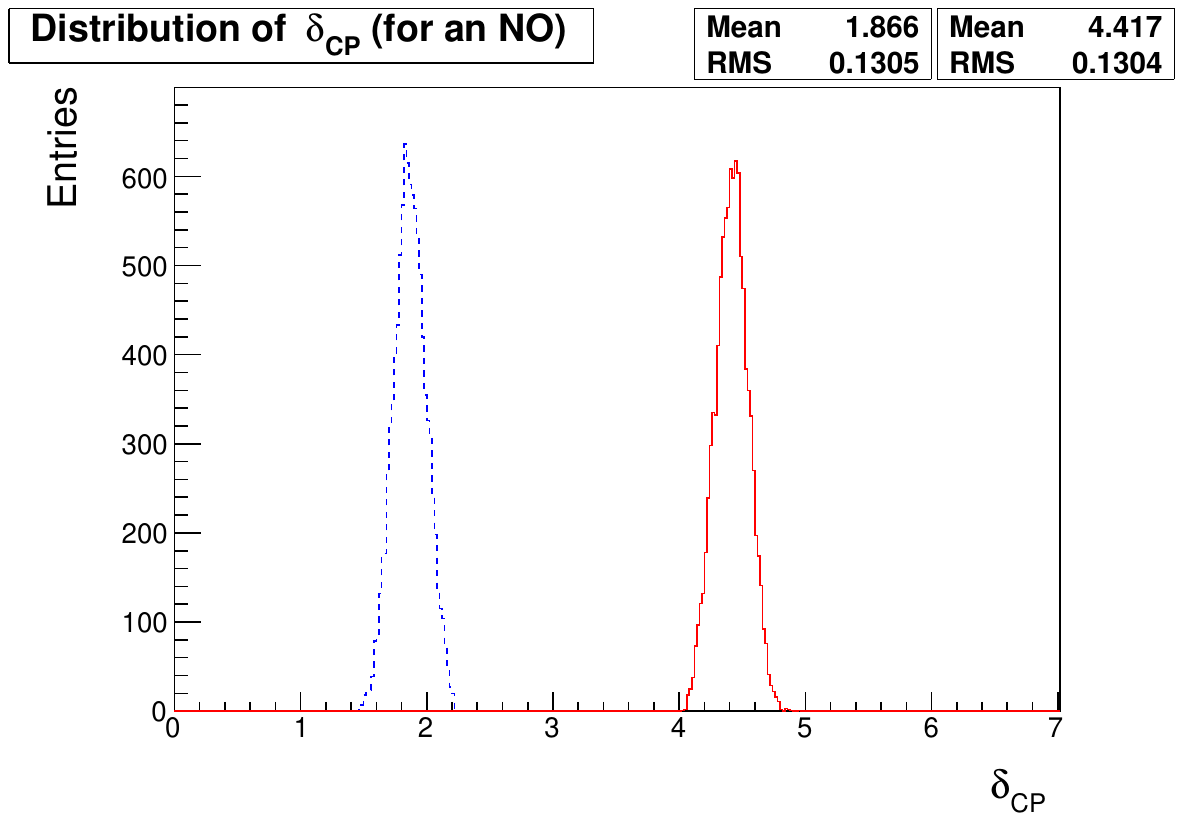}~~~&~~~
\includegraphics[width=6.5cm,height=5.5cm]{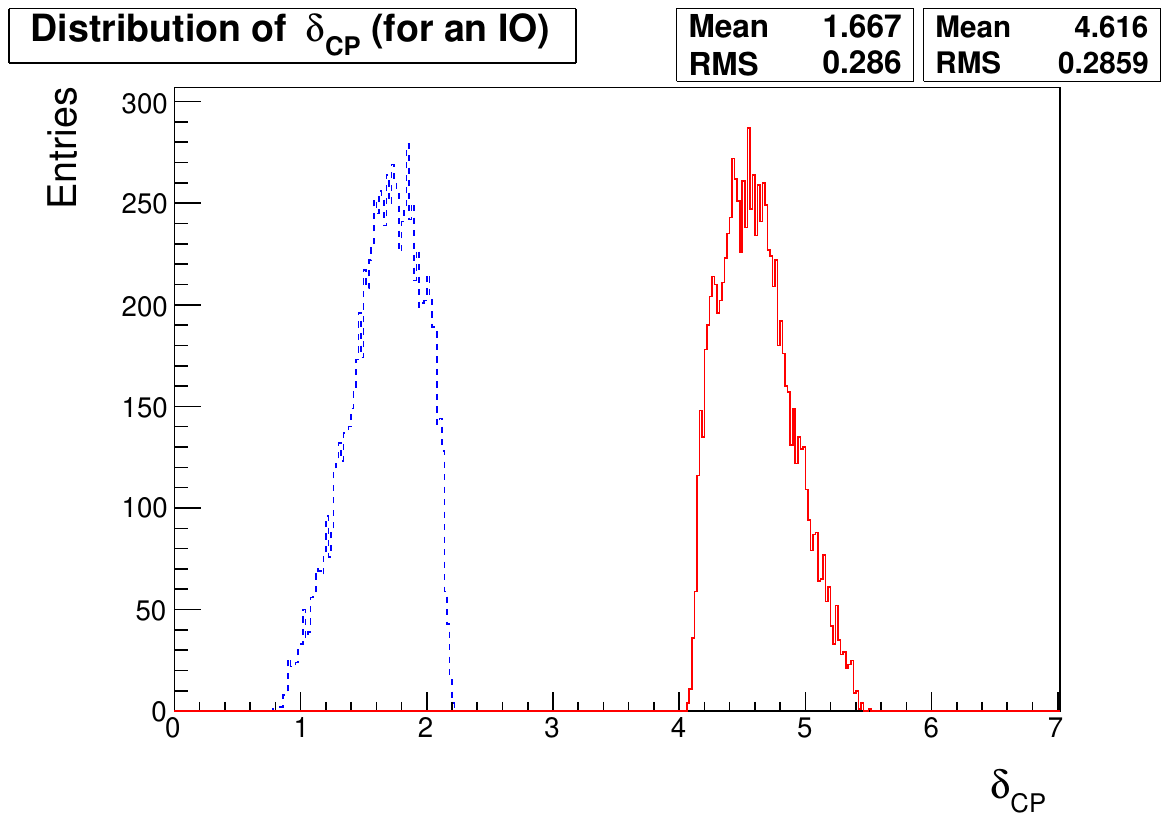}\\
\end{tabular}
\caption{\small Distributions of $\delta_{CP}$ in an NO and an IO for two 
solutions distinguished by red and blue.}
\label{Delta}
\end{center}
\end{figure}
\begin{figure}[H]
\begin{center}
\begin{tabular}{cc}
\includegraphics[width=6.5cm,height=5.5cm]{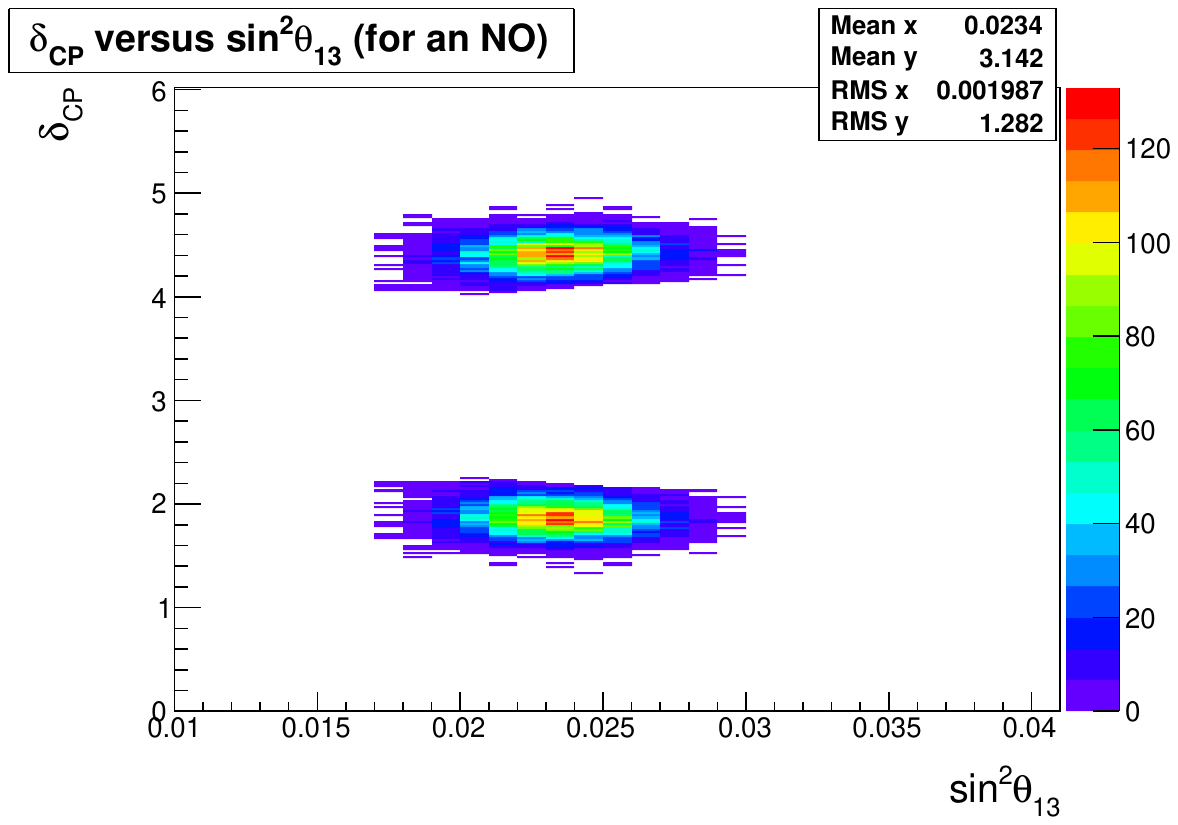}~~~&~~~
\includegraphics[width=6.5cm,height=5.5cm]{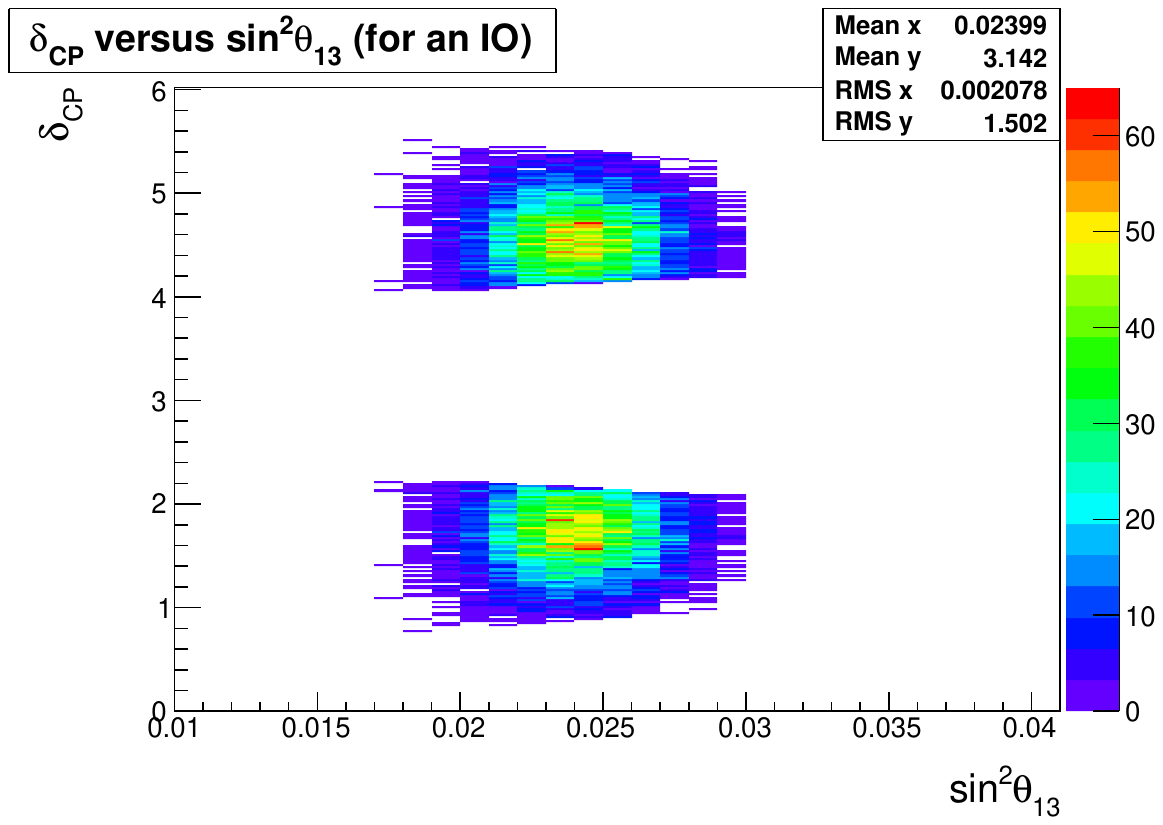}\\
\end{tabular}
\caption{\small $\delta_{CP}$ as a function of $\theta_{13}$ in an NO and an IO, where 
the 1$\sigma$, 2$\sigma$ and 3$\sigma$ regions are colored in red, green and blue, 
respectively.}
\label{Delta-s13}
\end{center}
\end{figure}
\end{center}
\end{widetext}

Using \eqref{J} and \eqref{Jmax}, the distributions of $J_{CP}$ corresponding to those 
of $\delta$ in two cases of hierarchy are depicted in Figs. \ref{Jcp} and \ref{Jcp-s13}. 
Because of the symmetry of $J_{CP}$ only its positive part is plotted in Fig. \ref{Jcp}. 
The values of $J_{CP}$ obtained around 0.032 (for an NO) and 0.034 (for an IO) are quite 
consistent with the ones given in \cite{Olive:2016xmw}.
\begin{widetext}
\begin{center}
\begin{figure}[H]
\begin{center}
\begin{tabular}{cc}
\includegraphics[width=6.5cm,height=5.5cm]{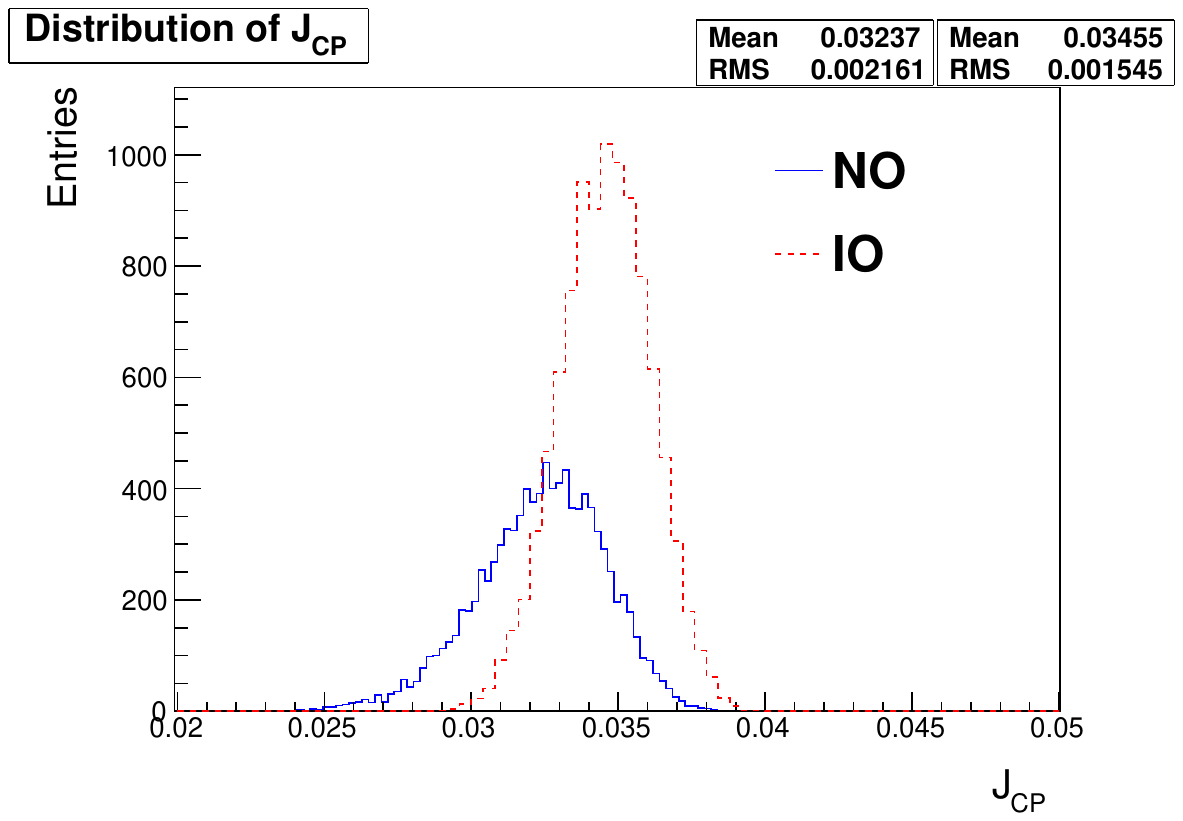} 
\end{tabular}
\caption{Distribution of $J_{CP}$ in an NO and an IO}
\label{Jcp}
\end{center}
\end{figure}
\begin{figure}[H]
\begin{center}
\begin{tabular}{cc}
\includegraphics[width=6.5cm,height=5.5cm]{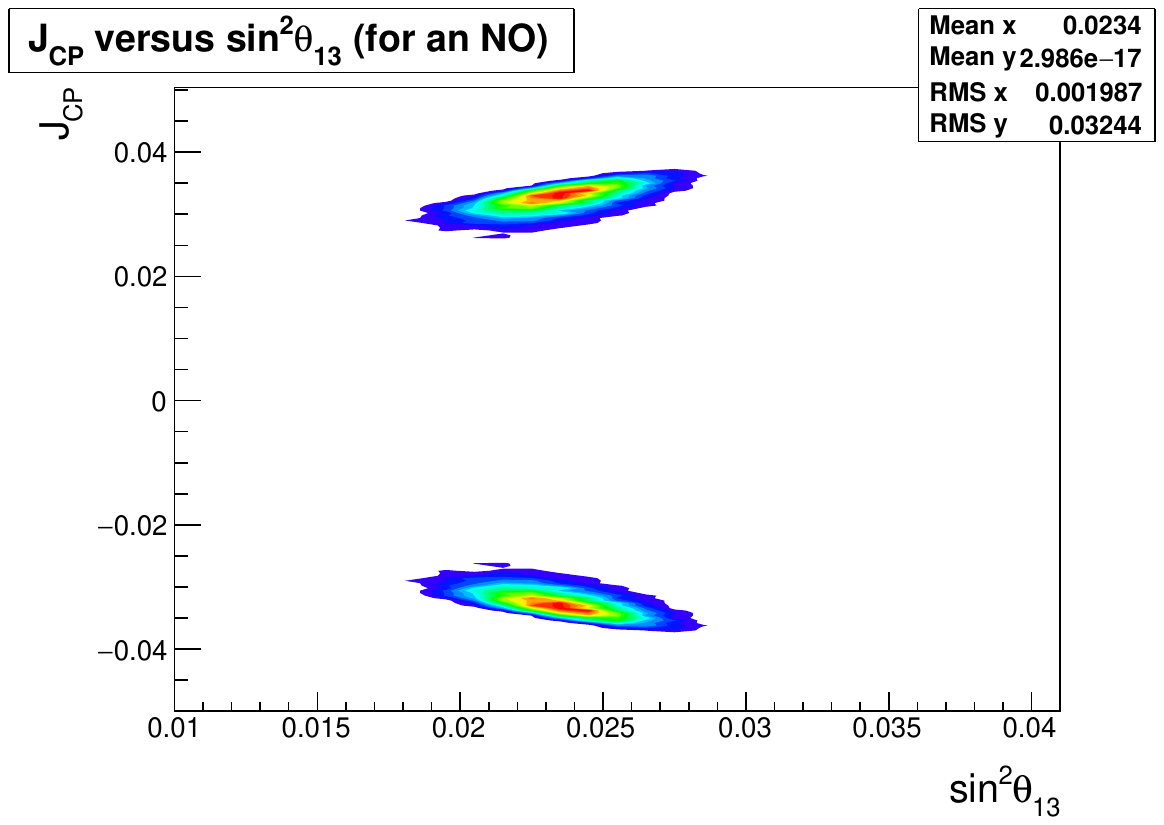}~~~&~~~
\includegraphics[width=6.5cm,height=5.5cm]{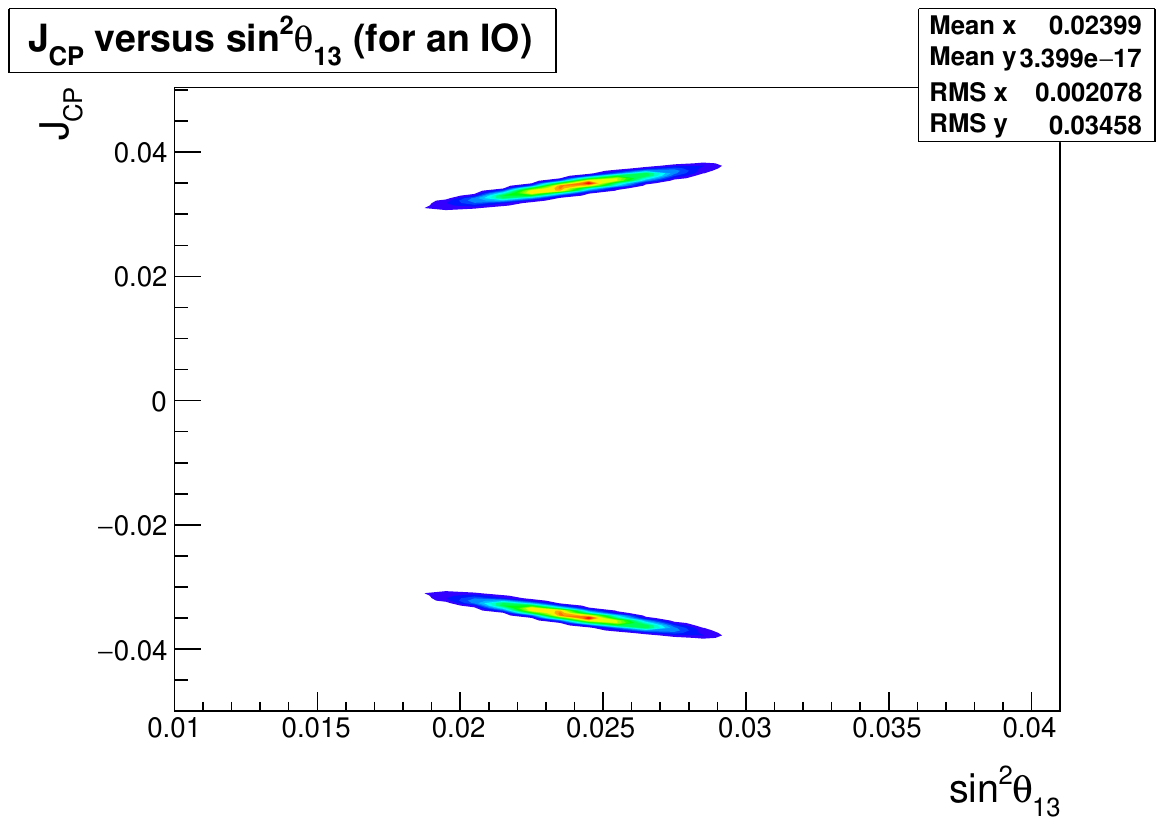}\\
\end{tabular}
\caption{\small $J_{CP}$ as a function of $\theta_{13}$ in an NO and an IO.}
\label{Jcp-s13}
\end{center}
\end{figure}
\end{center}
\end{widetext}
To summarise, we note that the best fit values of $\delta$ 
and $J_{CP}$ obtained here are very close (in general, at the $1\sigma$ 
region) to the global fit given in \cite{Olive:2016xmw,Capozzi:2017ipn, Esteban:2016qun}. 

\section{Conclusions}
\hspace{0.5cm}
We stress again that neutrino masses and mixing are an experimental fact. 
Neutrinos are massive but their masses are very tiny. Therefore, an actual problem is 
to find a way to explain the smallness of the neutrino masses and the 
see-saw mechanism is one of the most popular ways serving this purpose. So far 
many models of neutrino masses and mixing including those with the flavour 
symmetry have been proposed. In particular, the models based on the discrete 
group $A_4$ has attracted much interest and has been investigated quite intensively 
for the last several years. A natural question arising is whether and how the 
see-saw mechanism can be extended to the case of a flavour symmetric SM. 
The see-saw mechanism has been applied by other authors to separate 
sub-processes (classified by an $A_4$ symmetry), 
but, to our knowledge, no universal and systematic formulation of the see-saw 
mechanism has been made for an $A_4$-flavour symmetric SM yet. In the present 
paper such a formulation with an accent on a type-I see-saw model has been 
made. So, an ordinary see-saw process is treated as an effective process suming up 
all possible sub-processes corresponding to irreducible representations of the $A_4$ 
group.\\

The scalar sector of the latter model contains an $A_4$-triplet $\Phi_h$ of 
iso-doublet scalars and three $A_4$-singlet iso-singlet scalars $\Phi_S$, 
$\Phi_S^{'}$ and $\Phi_S^{''}$. In our scheme, these $A_4$-singlets when acquiring 
a VEV, violate the extended symmetry down to the SM one, the $A_4$-triplet would 
play the role of the SM Higgs. The real parts of the three neutral components of 
$\Phi_h$ are super-positions of three mass-eigen states, one of which could be the 
SM Higgs. A detailed analysis on the scalar sector is a subject of a separate work. \\ 

The lepton sector of the above-mentioned model consists of the SM-like leptons 
transforming now under also the group $A_4$ as its three- or one-dimensional 
representations, and four new iso-singlet fields transforming as an $A_4$-triplet 
and three $A_4$-singlets $1$, $1'$ and $1''$. These newly introduced 
iso-singlet fields, referred to as right-handed neutrinos, are necessary for the 
neutrino mass generation via the type-I see-saw mechanism. Applying a perturbation 
method to this model we obtain a neutrino mass and mixing structure from where 
several important quantities, such as neutrinoless double beta decay effective mass 
$|\langle m_{ee}\rangle|$, CP violation phase $\delta_{CP}$ and Jarlskog parameter 
$J_{CP}$ are predicted. The predicted values of these quantities within the present 
model fit quite well to the current experimental data \cite{Olive:2016xmw}. After finishing this work we have learned that our results are consistent to the results announced recently by the T2K collaboration \cite{Iwamoto:2016yth}, in particular, the values of $\delta_{CP}$ obtained by us, in both the NO and the IO, are very close to those announced by the T2K.\\

For conclusion, it is shown once again in this paper the usefulness of the see-saw 
mechanism applied to $A_4$ flavour symmetric standard models predicting new particles,  
interesting relations and values of some physics quantities which can be tested experimentally. Compared with our previous work \cite{Ky:2016rzl}, 
the advantage of the present approach is that no introduction of an additional 
symmetry $Z_3\times Z_4$ is needed. An illustration has been done within a type-I 
(see-saw) model, but it can be also done with other-type models, or with all-type 
models.\\    

\begin{acknowledgments}
This work is supported by the National Foundation for Science and Technology Development (NAFOSTED) of Viet Nam under the grant No 103.01-2014.89. 
\\

Three of us (D.N.D, N.A.K. and N.T.H.V.) would like to thank K. Narain for warm 
hospitality in the Abdus Salam ICTP, Trieste, Italy. N.A.K. would like to thank 
W. Lerche and L. Alvarez-Gaume for warm hospitality at CERN, Geneva, 
Switzerland, and Y. Sakai, S. Uno and M. Yamauchi for warm hospitality at KEK, Tsukuba, Japan. 
\end{acknowledgments}

\appendix
\section{Representations of $A_4$ in brief}
\label{append}
\mathversion{bold}
\mathversion{normal}
\hspace{0.5cm}
Being a group of all even permutations of four objects, $A_4$ has 12 elements, 
which can be divided into 4 conjugate classes, thus is also an order 4 group 
\cite{Ishimori:2010au, Altarelli:2010gt,Altarelli:2009kr}. 
Geometrically, it is also called the tetrahedral group describing the 
orientation-preserving symmetry of a regular tetrahedron. The elements of $A_4$ 
can be generated by two basic generators $S$ and $T$ satisfying the relations 
\begin{equation}
S^2=T^3=(ST)^3=1.
\end{equation}
This group has four unitary representations, including three one-dimensional 
representations $1,~1',~1''$ and one three-dimensional representation generated 
via $S$ and $T$ as follows:
\begin{subequations}
	\begin{equation}
	1: \hspace{0.2cm} S=1, \hspace{0.3cm} T=1,
	\end{equation}
	\begin{equation}
	1^{'}:\hspace{0.2cm} S=1, \hspace{0.3cm} T=e^{i2 \pi/3} \equiv \omega, 
	\end{equation}
	\begin{equation}
	1^{''}: \hspace{0.2cm} S=1, \hspace{0.3cm} T=e^{i4 \pi/3} \equiv \omega^2,
	\end{equation}
	\begin{equation}
	3: \hspace{0.2cm} S = \left(
	\begin{array}{ccc}
	1 & 0 & 0 \\
	0 & -1 & 0 \\
	0 & 0 & -1
	\end{array}
	\right),
	\hspace{0.3cm}
	T = \left(
	\begin{array}{ccc}
	0 & 1 & 0 \\
	0 & 0 & 1 \\
	1 & 0 & 0
	\end{array}
	\right).
	\end{equation}
\end{subequations}
Usually, applications of representations of a group require us to know the 
decomposition rule of a tensor product between irreducible representations 
into irreducible representations. This rule in the case of $A_4$ reads 
\begin{subequations}
	\begin{equation}
	1\times 1 = 1,
	\end{equation}
	\begin{equation}
	1^{'} \times 1^{''} = 1,
	\end{equation}
	\begin{equation}
	1^{''} \times 1^{'} = 1,
	\end{equation}
	\begin{equation}
	1^{'} \times 1^{'} = 1^{''},
	\end{equation}
	\begin{equation}
	1^{''} \times 1^{''} = 1^{'},
	\end{equation}
	\begin{equation}
	3 \times 3 = 1+1^{'}+1^{''} +3_a+3_s.
	\label{tensor33}
	\end{equation}
	\label{tenprod1}
\end{subequations}
The first five rules are trivial but the last one needs more explanation. 
For two triplets $3_A \sim (a_1,a_2,a_3)$ and $3_B \sim (b_1,b_2,b_3)$,  
the irreducible components of their tensor product $3_A\times 3_B$ according to (\ref{tensor33}) are 
\begin{subequations}
	\begin{equation}
	1 = a_1b_1+a_2b_2+a_3b_3,
	\end{equation}
	\begin{equation}
	1^{'} = a_1b_1+\omega^2 a_2b_2+\omega a_3b_3,
	\end{equation}
	\begin{equation}
	1^{''} = a_1b_1+\omega a_2b_2+\omega^2 a_3b_3,
	\end{equation}
	\begin{equation}
	3_a \sim (a_2b_3,a_3b_1,a_1b_2),
	\end{equation}
	\begin{equation}
	3_s \sim (a_3b_2,a_1b_3,a_2b_1).
	\end{equation}
	\label{tenprod2}
\end{subequations}
%

The information of the $A_4$ representations given above is important for 
construction of a Lagrangian of a model adopting an $A_4$ flavour symmetry.
Basing on the structure of tensor products 
of representations 
of $A_4$ we can build different see-saw models corresponding to this flavour symmetry.

\bibliography{apssamp}

\end{document}